# PRELIMINARY DESIGN STUDY OF HIGH-POWER H₂⁺ CYCLOTRONS FOR THE DAEδALUS EXPERIMENT

L. Calabretta, L. Celona, S. Gammino, D. Rifuggiato, G. Ciavola

*Istituto Nazionale di Fisica Nucleare, Laboratori Nazionali del Sud, Catania, I-95123, Italy*

M. Maggiore, L. A. C. Piazza

*Istituto Nazionale di Fisica Nucleare, Laboratori Nazionali del Sud, Catania, I-95123, Italy*

J. R. Alonso, W.A. Barletta, A. Calanna[1], J.M. Conrad

*Massachusetts Institute of Technology, Cambridge, MA 02139, USA*

(Dated: 2 July, 2011)

## ABSTRACT

This report provides a first design for H₂⁺ accelerators as the DAEδALUS neutrino sources. A description of all aspects of the system, from the ion source to the extracted beam, is provided. This analysis provides a first proof of principle of a full cyclotron system which can provide the necessary beam power for the CP violation search proposed by the DAEδALUS Collaboration.

---

[1] *Present address: Centro Siciliano di Fisica e Struttura della Materia, I-95123, Italy*

# 1. INTRODUCTION

The DAEδALUS (Decay At rest Experiment for $\delta_{cp}$ At Laboratory for Underground Science) Experiment is a new approach to search for CP violation in the neutrino sector [Alon2010a, Alon2010b]. The present project develops the design of a high power accelerator "module" capable of supplying a proton beam of ~800 MeV, 8 MW peak power, onto a graphite target as a source of neutrinos. The module will typically operate at a duty factor of ~20%, nominally planned as 100 msec beam on, 400 msec beam off, although this specification is flexible. At 20% duty factor the average power from the module is 1.6 MW.

The DAEδALUS experiment needs three independent "stations," each containing one or more accelerator modules, depending on the total beam power needed to obtain the required neutrino flux from the targets. This "complex" of stations, originally configured for deployment at the Sanford Laboratory (Lead, SD), would have the nearest station located at 1.5 km from (and directly above) the underground detector, and would have a minimum power of 1 MW. The second station would be at a distance of about 8 km from the detector and would need an average beam power of 2 MW. The last station, 20 km away from the detector, would be supplied with proton beams of average power of about 6 MW. Neutrinos produced by the three sources are detected by an ultra-large (~300 kT) water Cherenkov detector doped with Gd. The three sources have beam-on times synchronized, so the detector will receive the ~100 msec beam bunch from each source sequentially. For ~200 msec (40% of the total cycle time), all sources are off, to allow measurement of cosmogenic and other backgrounds. In this configuration, a full cycle would have a 500 msec period. This timing and arrangement of accelerator stations allows a precise search for CP violation using the $\bar{\nu}_\mu \to \bar{\nu}_e$ oscillation signals from each source.

Although the required average power for the first two stations is 1 and 2 MW respectively, the need to operate at ~20% duty factor has the consequence that substantially higher peak powers and beam currents are necessary. At the same time the lower beam average power mitigates the problems related to thermal dissipation and activation. The required currents are well into the region where space charge effects become extremely relevant both for the injection process and for extraction efficiency. Hence, solutions that mitigate the space charge effects are extremely important, and are the underlying rationale for our design.

Accelerator complexes consisting of two or three cyclotrons, one or more injector cyclotrons and a main ring cyclotron booster, have already been proposed as drivers for energy amplifiers or waste transmutation plants [Stam1996, Cala1999a, Jong1999]. The main requirements and constraints for such designs are: proton currents higher than 10 mA and energy as high as 1 GeV, minimum beam losses, high reliability and high conversion efficiency from electrical (so-called "wall" power) to beam power.

Well-known conventional cyclotron designs are quite well-suited as reliable and economical solutions for a plant which requires a peak beam power of 1-5 MW [Cala1999a, Jong1999]. To deliver higher peak power, i.e. 10 or more MW, some important problems for a ring cyclotron design must be addressed: space charge effects, extraction systems and power dissipation in each of the accelerating RF cavities. To overcome these problems the traditional solution is to increase the radius of the cyclotron and the number of cavities. But this significantly increases the plant cost.

An alternative solution based on the acceleration of molecular $H_2^+$ has been proposed [Cala1999b, Cala1999c]. In this case the extraction of the $H_2^+$ beam is accomplished by a stripping foil that dissociates the molecule, producing two free protons. Due to the different magnetic rigidity as compared with the $H_2^+$, the protons escape quite readily from the magnetic field of the cyclotron. Extraction by stripping does not require well-separated turns at the extraction radius and allows using lower energy-gain per turn during the acceleration process and/or lower radius for the magnetic sectors, with a significant reduction of thermal power losses for the RF cavities and all resulting in lower construction cost. Stripping extraction also allows for beams with large energy spread (0.5 – 1%). As a result, the energy spreading produced by the space charge effect on the longitudinal size of the beam is not crucial in this accelerator, and flat-topping cavities are unnecessary. We believe that the acceleration of an $H_2^+$ beam, despite the need to handle a beam with magnetic rigidity twice that of protons, offers a remarkable advantage in terms of reliability, ease of operations, and lower cost for both construction and operation.

Twelve years ago, a layout for an accelerator able to supply a proton beam with energy of 1 GeV and a beam power up to 10 MW was developed [Cala1999b]. The goal of that design was to drive a sub-critical reactor. That previous proposal has now been updated to fit the requirements of the DAEδALUS neutrino experiment. An important aspect of the experiment is that the number of accelerator modules required is at least three or four, placing a strong requirement on minimizing the cost per accelerator.

The solution presented here, shown schematically in Figure 1, shows a module consisting of two cascaded cyclotrons. The injector cyclotron, a four-sector machine, accelerates a beam of $H_2^+$ up to energy of about 50 MeV/n. The beam is then extracted by an electrostatic deflector and is transported and injected into an 8 sector Superconducting Ring Cyclotron. Two stripper foils are used to extract two proton beams at the same time from the ring cyclotron. This solution allows an increase of the mean life of the stripper foils and may reduce the issues in the design of the target/beam dump.

The peak current of the cyclotrons in this module is 5 mA of $H_2^+$ – 10 mA of protons after stripping – or 8 MW of peak power. If each module is operated at 20% duty factor, the average power is 1.6 MW per module.

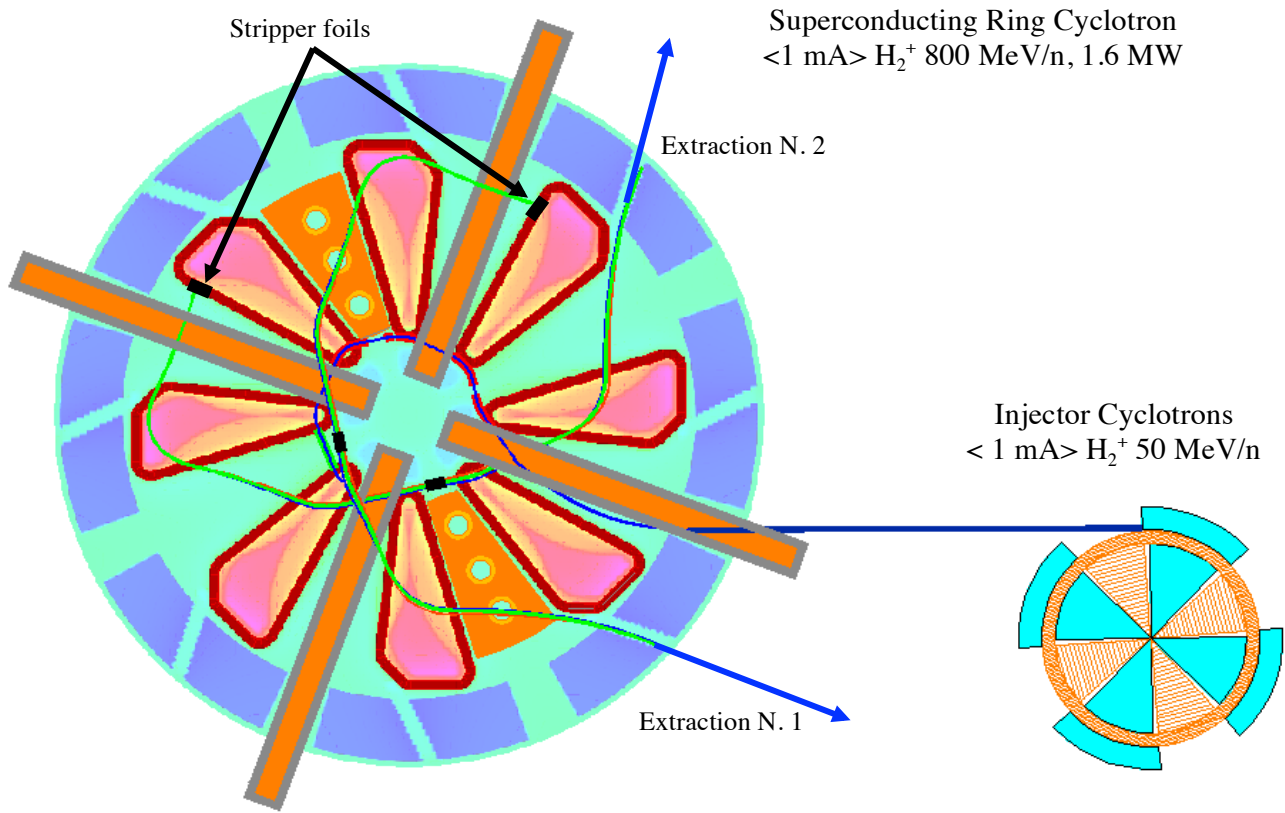

*Figure 1.1: Layout of the accelerator module, the average beam current <1 mA> is given for a 20% duty cycle.*

## 2. POSSIBLE CONFIGURATIONS FOR THE CYCLOTRON COMPLEX

The design of the DAEδALUS experiment calls for 3 neutrino sources fed by proton beam-powers of 1, 2 and 6 MW respectively at distance of 1.5 km, 8 km and 20 km.

The Superconducting Ring Cyclotron (SRC), in the configuration described above, will be able to supply a peak of 5 mA of $H_2^+$ (i.e., a peak particle current of 2 mA of protons) up to an energy of 800 MeV/n. When operated at a 20% duty factor, beam power per module will be 1.6 MW. The near and middle stations could each be equipped with one module with slight adjustments of the duty factors for each. The first module could be operated with a duty cycle of 12.5%, the second can deliver the 2 MW average beam power if we drive the module with a duty factor of 25% instead of the 20%. For the assumed repetition cycle of 500 msec, the first module will deliver beam for 62.5 msec, while the second module will deliver beam for 125 msec. Delivering 6 MW at the 3rd station is more challenging. The straightforward solution is to use 3 cyclotron modules operating in the same mode as station #2. In this case, the time period when the beam is "ON" in one of the 3 sites is 62.5+125+125 = 312.5 msec, while the time when the beam is "OFF" is reduced from the requested 40% or 200 msec, to 187.5 msec or 37.5%. This difference is not considered a serious problem for the neutrino oscillation studies.

To reduce the cost of the facility, we consider some alternatives for the configuration of the third station. The maximum beam power delivered by the cyclotron module described above is limited mainly by

- Maximum beam current delivered by the source of $H_2^+$. Today this is 20 mA. With future source optimization this could be increased to 40 mA or higher;
- Maximum current delivered by the injector cyclotron as a consequence of the space charge effects which are dominant at low velocity, that can limit extraction efficiency from this first cyclotron, and would be worse for higher currents from the ion-source;
- Beam losses at high energies, expected to be primarily due to the interaction with the residual gases inside the Superconducting Ring Cyclotron.

Note that the maximum beam power delivered by the cyclotron module is limited by the current delivered by the injector cyclotron, which in turn is determined mainly by the space charge effects that are dominant at injection. Space charge forces can increase the beam emittance and consequently reduce the extraction efficiency of the cyclotron injector. Reduced efficiency could be a serious limitation because beam losses must be kept below 200 W for each cyclotron vault. This limit, following the experience of PSI and other laboratories, is found to be a practical maximum to allow for routine hands-on maintenance in the cyclotron vault. Note, at 100 MeV total energy, 200 watts corresponds to an allowed beam loss of 2 $\mu$A of $H_2^+$, or two parts in $10^3$. One must pay close attention to extract the beam cleanly from the injector cyclotron.

In contrast, space charge effects are less relevant in the ring cyclotron, because of stripping extraction. However, residual gas interactions could dominate the beam-loss budget. We are planning for a vacuum design goal of at least $1 \times 10^{-8}$ Torr (see Section 10); simulations indicate this pressure will keep gas-interaction beam losses to an acceptable level. Lower pressures would allow for acceleration of even higher beam currents without experiencing gas-interaction losses.

If space-charge effects can be controlled, simulations indicate that substantially higher peak currents could be extracted from the injector cyclotron, perhaps as high as 10 mA or more. This, however, would place enhanced requirements on the source current and brightness.

Two options may allow increasing the power-output per accelerator module:

1. Without increasing the injector cyclotron maximum average current of 1 mA, merging beams from two identical cyclotron injectors could double the $H_2^+$ current injected into the Ring Cyclotron. A novel merging scheme has been proposed [Owen2011], based on extracting beam from the two injector cyclotrons at slightly different energies, and merging them after passing through a dipole then removing the energy difference with a small RF cavity. This scheme is shown in Fig. 2.1. In this case the RF frequency of the injector cyclotrons must be lowered to 24.58 MHz (half the 49.16 MHz of the SRC) to allow merging of alternate beam bunches.

2. If the injector cyclotron is able to extract up to 2 mA of $H_2^+$ average current (peak current 10 mA), then a doubling of the ion source current, and mitigating the central-region space-charge effects, possibly by a higher-energy injection scheme, could allow a single injector cyclotron to provide all the necessary current. There is substantial confidence that developing an ion source with twice the $H_2^+$ current will be within technical capabilities. Higher-energy injection might be achieved with an RFQ (radio-frequency quadrupole) pre-injector, a well-established technology, but not one easily matched to a cyclotron. However, there is an existence proof, from the Hahn-Meitner VICKSI accelerator facility, so the issues of longitudinal emittance and bunch length matching might be successfully addressed.

If either of these options can be realized, the ring cyclotron should be able to deliver a beam power of 16 MW in CW mode or 3.2 MW with a 20% duty factor. In this way, the third DAEδALUS station power requirement could be met by installation of two such "super" modules, supplying a total power of 6.4 MW.

The configurations described here allow delivery of different amounts of beam power at the different sites using similar cyclotron modules. This offers very great advantages in cost-reductions for design and construction, as well as for increased efficiency from commonality in operation and maintenance procedures.

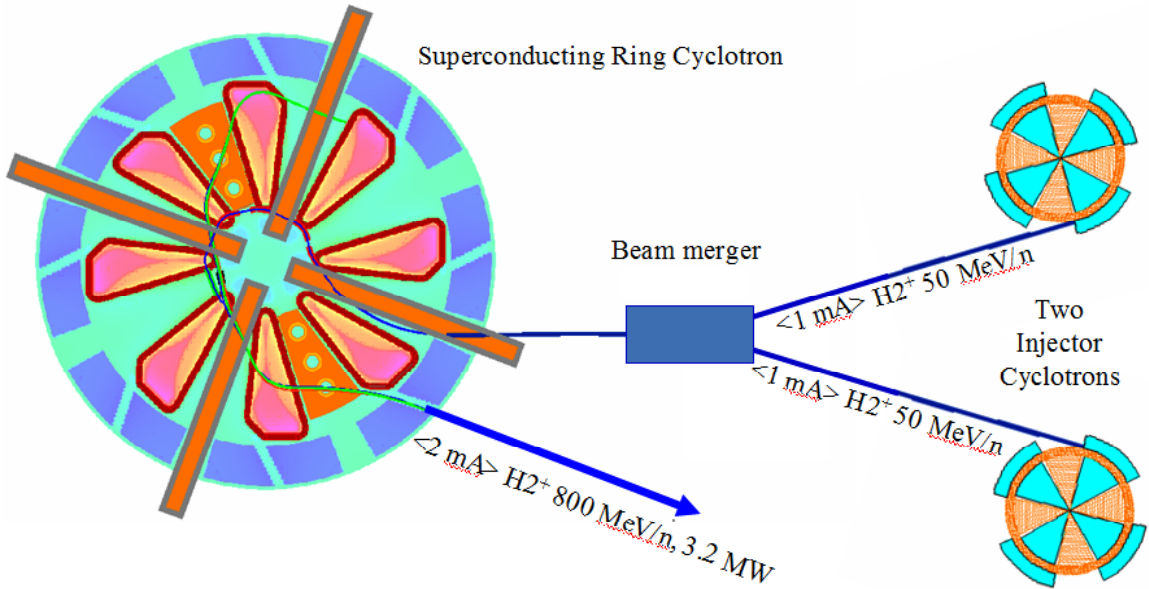

*Figure 2.1: Possible layout of a cyclotron "super" module for the station at 20 km. Alternate RF bunches are merged into a single train injected into the SRC. The two cyclotron injectors are driven at the 3$^{rd}$ harmonic (24.58 MHz), while the ring cyclotron is operated at the 6$^{th}$ (49.16 MHz). Beam merger is described in [Owen2011].*

## 3. SPACE CHARGE EFFECTS

Single-particle beam dynamics assumes negligible interactions between the particles of the beam. This assumption is not valid when the beam current exceeds 1 mA. Here, we compare the space charge effects for $H_2^+$ vs. a proton beam. The space charge of the particle beam produces a repulsive force inside the beam bunches, thereby generating detuning effects. A measure of the strength of these effects is called the generalized perveance, [Reis2008] defined by the following formula:

$$K = \frac{qI}{2 \cdot \pi \cdot \varepsilon_o \cdot m \cdot \gamma^3 \beta^3} \quad (3.1)$$

where $q$, $I$, $m$, $\gamma$ and $\beta$ are respectively the charge, current, mass and the relativistic parameters of the particle beam. The higher the value of K, the stronger the detuning effects. Formula (3.1) implies that the proton beam has a perveance double that of the $H_2^+$ beam of the same velocity. However, if protons and $H_2^+$ are accelerated by the same electric field, they have the same energy but not the same $\beta$. On the other hand, a beam of $H_2^+$ delivers twice the number of protons as a proton beam with the same electrical current. Therefore a more appropriate comparison is for beams carrying the same proton current with the same total energy or same beam velocity.

Table 3.1 compares the perveance of $H_2^+$ and proton beams with a current of 5 mA and 10 mA, respectively, at the same and at different energies. The ratio of perveance values shows that, with respect to the space charge effects, accelerating a $H_2^+$ beam is less difficult than accelerating a proton beam with double the current. This advantage increases with higher beam energy.

The last two rows of Table 3.1 show the perveance values of a proton beam with a current of 2 mA and the ratio vs. the perveance of a $H_2^+$ beam with 5 mA. Although the perveance of the 2 mA proton beam is lower than the same-energy $H_2^+$ beam (30 keV), if the energy of the $H_2^+$ beam is increased up to 70 keV, then the $H_2^+$ beam has the same perveance of the proton beam at 30 keV. Therefore a $H_2^+$ beam with an energy of 70 keV and with current of 5 mA suffers the same space charge effects as a proton beam with energy of 30 keV and 2 mA.

Present commercial cyclotrons for radioisotope production can deliver proton beams at 30 MeV with a beam current of 1.6–2 mA. Though these all accelerate $H^-$ and use stripping extraction, the central-region capture and acceleration characteristics will be similar to our requirements. Consequently a cyclotron for $H_2^+$ with focusing properties and beam dynamics similar to that of commercial cyclotrons should provide a good starting point for our design. The main requirements are to use an ion source able to deliver $H_2^+$ at energy as high as 70 keV (35 keV/n) and to accelerate the $H_2^+$ beam using an RF voltage approximately double that used in commercial cyclotrons.

Table 3.1: Perveance values of proton and $H_2^+$ beams at various energies.

|  | $E_p=E_{H2}$ @ 30 keV | $E_p=E_{H2}$ @ 800 MeV | $E_p$=30 keV $E_{H2}$=70 keV |
|---|---|---|---|
| $H_2^+$, I=5 mA | $0.881\ 10^{-3}$ | $0.151\ 10^{-9}$ | $0.247\ 10^{-3}$ |
| P, I=10 mA | $1.245\ 10^{-3}$ | $1.075\ 10^{-9}$ | $1.245\ 10^{-3}$ |
| $K_{H2}/K_p$ | 0.707 | 0.141 | 0.198 |
| P, I=2 mA | $2.491\ 10^{-4}$ | $2.15\ 10^{-10}$ | $2.491\ 10^{-4}$ |
| $K_{H2}/K_p$ | 3.537 | 0.703 | 0.992 |

## 4. INJECTOR CYCLOTRON

The injector II of the PSI cyclotron and the commercial compact cyclotrons designed by EBCO and IBA are presently the only cyclotrons that deliver more than 1.5 mA of proton beam. The injector II of PSI is a conservative design that can supply up to 3 mA of protons. It is a separated sector design with beam injection at 800 keV, final energy of 72 MeV, energy gain per turn of 1 MeV, extraction radius of 3.3 m, and single turn extraction using an electrostatic deflector. Despite low energy injection (25-30 keV) of the IBA and EBCO machines, and moderate energy gain per turn (<200 keV/turn), the compact commercial cyclotrons are able to accelerate $H^-$ beams with a current of 1.5-2.2 mA [Tsut2010]. These accelerators use a stripper foil to extract to the beam, and so are unsuitable for an $H_2^+$ injector.

Because the perveance of an $H_2^+$ beam with a current of 5 mA and with energy of 70 keV is similar to the perveance of a proton beam with 2 mA and energy of 30 keV, we propose a design which is a combination of the PSI injector II and of the compact commercial cyclotron cited above. The central region of the proposed injector is a scaled up central region of the commercial cyclotron. To account for the higher magnetic rigidity and to maintain the perveance of the $H_2^+$ beam similar to that of the proton beam injected into the commercial cyclotrons, both the total injection energy and the energy gain per turn must be doubled. In fact, the energy gain per turn should increase radially up to 2 MeV at the extraction radius, a value higher than the energy gain per turn in the PSI injector. This choice compensates for the smaller extraction radius and for the lower charge to mass ratio (q/A=0.5 in our proposal vs. the q/A=1 of the PSI injector.)

The parameters of the injector cyclotron are strictly correlated with the parameters of the ring cyclotron. In particular the operating RF frequency of the injector cyclotron must be a sub-harmonic of or equal to that of the ring cyclotron. The most recent version of the DAEδALUS ring cyclotron operates at an RF frequency of 49.16 MHz which is the 6[th] harmonic of the natural revolution frequency inside the ring cyclotron. The injection radius in the ring cyclotron is about 1.85 m. Therefore, many of the parameters for the injector cyclotron are constrained according to the parameters presented in Table 4.1. The average magnetic field at extraction, "<B> at $R_{ext}$", is equal to the $B_o$ at the injection orbit of the ring cyclotron. Consequently, the revolution frequency is the same for both cyclotrons. The operating frequency of the RF cavities of the injector cyclotron is the 6[th] harmonic, which is exactly the RF frequency of the ring cyclotron.

Table 4.1: Parameters of the Injector Cyclotron

| Einj | 35 keV/n | Emax | 50 MeV/n |
|---|---|---|---|
| Rinj | 50.1 mm | Rext | 1.85 m |
| <B> at Rinj | 1.075 T | <B> at Rext | 1.133 T |
| Sectors | 4 | Cavities | 4 |
| Cavity width | 37° | Cavity type | Double gap |
| RF | 49.16 MHz | Harmonic | 6[th] |
| V @ Rinj | >70 kV | V @ Rext | 280 kV |
| ΔR at Rext | 17 mm | ΔE/turn | 2.0 MeV |
| Beam width | <15 mm | Turns | < 85 |
| Peak current | 5 mA | Duty factor | 20% |
| <I> of $H_2^+$ | 1 mA | <Beam power> | 100 kW |
| Injection eff. | 10 - 15 % | Extraction eff. | > 99.8% |
| Extraction: Electrostatic Deflector + Magnetic Channels | | | |
| Deflector Gap | 20 mm | Electric field | <30 kV/cm |

While matching isochronism in the magnetic field is essential, the advantage from maximizing the energy gain per turn leads to using a small width for the cyclotron valley[*], about 39°. The width of the hill is assumed to be 51° so as to achieve the required average field of 1.075 T – assuming a maximum magnetic field for the fully saturated pole of 1.9 T.

Detailed simulation of beam dynamics including the space charge effects are needed to clarify whether the extraction efficiency can meet the required value > 99.8%. The acceptable maximum beam losses throughout the acceleration and extraction processes must be less than 200 W, which is 0.2% of the required beam power from the injector. Note, the 200 W figure relates to beam loss above the Coulomb barrier that leads to component activation. Power lost at lower energies requires good cooling for the slits and collimators on which the beam is deposited, but does not contribute to activation. If the injection efficiency is 10%, and injection energy is 70 keV (total energy), approximately 1 to 2 kW will be lost during capture of the beam.

---

[*] The vertical field in an isochronous cyclotron varies azimuthally in a succession of high and low field regions. The high field regions are have narrow gaps between the pole pieces and are called hills; the low field regions have wide gaps and are called valleys.

We expect the beam simulation of the PSI injector II cyclotron to be representative of our design. Although the extraction radius of the proposed DAEδALUS injector cyclotron is smaller than that of the PSI injector, the extraction efficiency should be similar. Indeed the ratio of the voltage at extraction to voltage at injection of the present design is 3.57 versus a value of 2 for the PSI injector. This higher voltage slope should produce stronger longitudinal focusing or bunch compression. The experience with the PSI injector cyclotron and the simulations with PSI's OPAL code [Yang2010][Bi__2011], including the space charge effects, indicate that the extraction efficiency might even be better for our beam species, currents and energies.

Although the final turn separation at extraction in our injector is 17 mm, compared with ~23 mm for the PSI injector II, beam losses at extraction should be kept at the required < 0.2%, using the precession of the beam orbit in the extraction region. Thus, expected beam losses at the extraction should be < 200 W for a delivered average beam power of 100 kW. The beam power extracted from PSI injector II is now about 160 kW, and the beam lost at the extraction are roughly 0.1%. The much lower energy injection in our design, 70 keV vs. 800 keV for the PSI injector, requires the ion source to deliver a higher brightness beam, with smaller transverse and longitudinal emittances. But as seen below, the inevitable beam losses in the capture and bunching process yield much less power deposition on the central-region collimators than in the PSI injector.

As a comparison with a commercial cyclotron, Table 4.2 gives the main parameters of the C-30 cyclotron of the IBA Company. Figure 4.1 shows the central region for the IBA C30, which should be quite similar to the central region of the DAEδALUS injector. The positions of the accelerating gaps used to define the radial beam size are indicated. We expect that the average beam power to be dissipated on each collimator in the DAEδALUS injector will be about 300 W. This value assumes the use of a beam buncher in the axial injection line to enhance the longitudinal acceptance. In this condition we expect the injection efficiency will be at least 15%, in a phase acceptance range of ±10° around the reference phase.

Table 4.2: Parameters of C30 Cyclotron, IBA Company

| | | | |
|---|---|---|---|
| Einj | 30 keV | Emax | 30 MeV |
| Rinj | 30 mm | Rext | 750 mm |
| <B> at Rinj | 1.0 T | <B> at Rext | 1.3 T |
| Sectors | 4 | Cavities | 2 |
| Cavities | λ/2 | Cavities | Double gap |
| RF | 66 MHz | Harmonic | 4$^{th}$ |
| ΔE/turn | 170 keV | Imax | 2 mA |
| Ion source current, H- | 15 mA | Injection efficiency | 13% |

The duty factor required by the DAEδALUS experiment is just 20% for each accelerator module. It is not clear yet what the optimum way of generating this duty factor is, whether to use long running periods (hours or days) at 100% on-time followed by longer periods when the machine is off, or to introduce methods of suppressing the beam 80% of the time while keeping the rest of the accelerator systems operating.

A chopper in the inflection line could be used, but the chopper must absorb as much as 2 kW of beam power. At 15% efficiency, and 5 mA peak current, source current must be 34 mA; at 70 keV total energy, this is 2.4 kW of beam power. If 80% is dumped into a block, this is close to 2 kW lost.

Although 34 mA is more than the currently-demonstrated 20 mA, it is not expected that doubling the $H_2^+$ current will present a significant problem. Proton sources in the 100 mA range exist; getting 40 mA of $H_2^+$ should be straightforward.

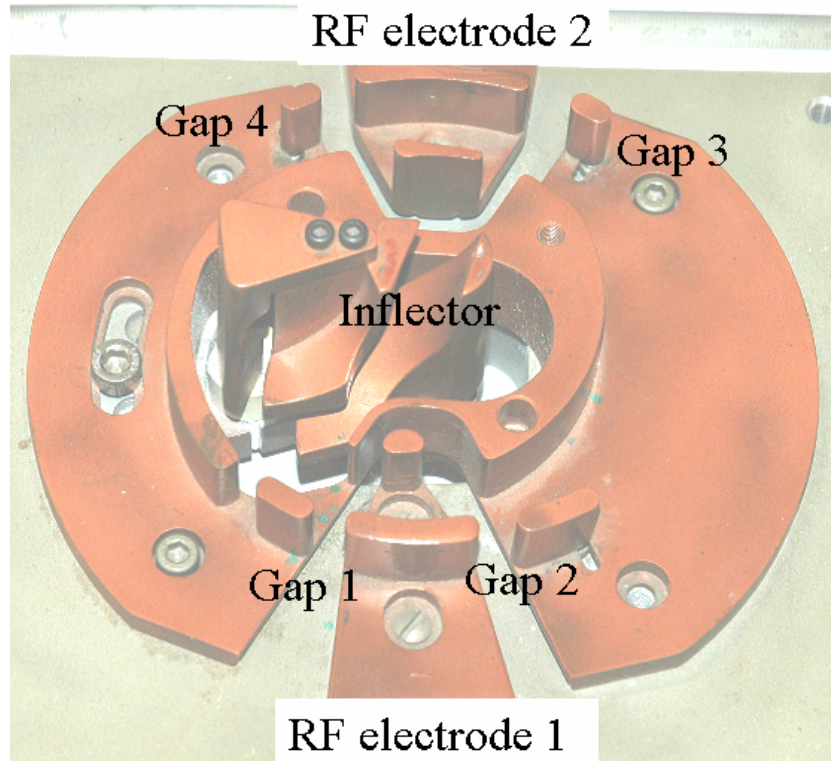

*Figure 4.1: Central region of C30 Cyclotron of IBA Company. The inflector, the RF electrodes and the accelerating gaps are shown.*

A more elegant option is to pulse the ion source, either by driving the microwave generator, if this type of source is used, or pulsing the HV for the arc plasma for a cusp source, with a pulse-shape that fits the DAEδALUS duty cycle. About half of the 6.8 mA (20% of 34 mA) from the source will be lost on the first collimator, or about 250 W.

In the region of the second gap a second collimator collects an additional 1.5 mA. At this position the average energy of the beam is ~210 keV (assuming an energy gain of 70 keV at each acceleration gap), and the beam power to be dissipated is about 300 W. Similarly, on the third collimator placed after the third acceleration gap, the current to be suppressed is about 1.5 mA with an average energy of 280 keV and a power loss of 400 W. The remaining 1 mA is accelerated to full energy.

Although the present estimate of beam loss is rough, it suffices to indicate that the beam power to be dissipated on each collimator is less than 400 W. This power is significantly lower than the beam power deposited on the collimators of the PSI injector (about 8 kW) and is similar to the beam power lost in the central region of the C30 commercial cyclotron. The beam losses on the collimators of the central region are insignificant in term of activation of the cyclotron and its vault, because the beam energy is less than 1 MeV.

# 5. $H_2^+$ ION SOURCE

Due to the low efficiency at injection accelerating a beam with a peak current of 5 mA (or 1 mA with 20% duty cycle), requires a source of $H_2^+$ able to deliver a beam current of 35 to 40 mA. A parasitic beam of $H_2^+$ is always produced in any kind of proton source. Optimization of a multicusp filament source for $H_2^+$ production was reported by Ehlers and Leung [Ehle1983]. This source delivered over 80% of the extracted beam in the $H_2^+$ fraction, with a reported current density of 50 mA/cm$^2$. Recently, LNS in Catania [Gamm2010] built a compact ECR, the Versatile Ion Source (VIS), able to deliver up to 33 mA of proton beam. Tests of the VIS show a parasitic beam of $H_2^+$ as large as 20% of the proton beam. Optimisation of the source characteristics, such as position of the permanent magnets, vacuum pressure, RF power, should allow beam currents of $H_2^+$ exceeding 20 mA. During a recent test, a beam current of 20 mA of $H_2^+$ has been measured for this source [Gamm2011]. Increasing the size of the source's extraction hole should sufficiently increase the beam current at the price of increasing the beam emittance. Other two important characteristics of the VIS are its good normalised beam emittance, ~0.1 π mm.mrad, and its extraction voltage which could be raised up to 70 kV. Both these two parameters fit very well with the requirements of the injector cyclotron, and may increase the injection efficiency.

Higher current sources have been developed for the IFMIF (International Fusion Materials Irradiation Facility) project, either microwave type [Gobi2008] or volume discharge multi-cusp [Holl2000]. These sources are designed to supply more than 100 mA CW of deuteron beams at 100 keV, with a normalized emittance of about 0.3 π mm.mrad or better. Although IFMIF is nominally a source for deuterons, initial tests to optimize the source optics will be performed with the production of $H_2^+$ beams. As $H_2^+$ is essentially identical to the mono-atomic deuteron beam with respect to the beam dynamics and space charge effects, extraction and transport optics should not change from the IFMIF parameters. Hence, these ion sources certainly could supply the $H_2^+$ beam for the injector cyclotron.

So, several source options exist for DAEδALUS, all likely to produce adequately bright beams with suitable currents and emittances. As will be described in Section 10.1, $H_2^+$ ions are produced in a wide range of long-lived vibrational states, ultimate source selection will depend on the ability to quench loosely-bound states that are likely to dissociate in the high SRC fields.

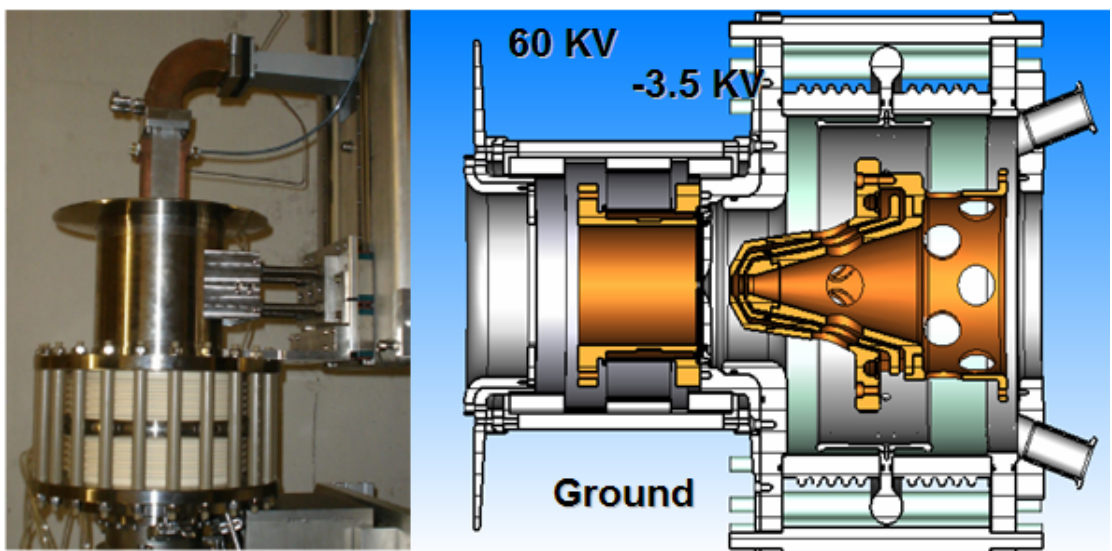

*Figure 5.1: Picture and layout of the Versatile Ion Source (VIS) developed at INFN-LNS, Catania*

# 6. SUPERCONDUCTING RING CYCLOTRON

The main component of the accelerator complex is the Superconducting Ring Cyclotron (SRC). It consists of 8 sectors of superconducting magnets that produce a magnetic field satisfying the required cyclotron isochronicity condition[†]. With proper focusing properties, the acceleration and the extraction of the beam are guaranteed. The $H_2^+$ beam will be injected into the ring along one of the cyclotron valleys using one or more superconducting injector magnets and one electrostatic deflector. Acceleration will be performed using at least 6 RF cavities. Beam extraction is performed by insertion of a pyrolitic graphite stripper foil with thickness less than 2 mg/cm$^2$ (10 micrometers). It is possible to use one or two stripper foils consistent with the power limitations of the neutrino-target/beam-dump. Although the present study requires further refinement, the results achieved so far satisfy all DAEδALUS physics requirements.

## 6.1 The Sector Magnet for the Superconducting Ring Cyclotron

To date, we have simulated about 200 different models of the sector magnet. The goal of the optimization process was two-fold: 1) produce a configuration of the magnetic sector that generates the isochronous magnetic field with good focusing properties in both the radial and vertical planes and with a reliable shape; 2) minimize the current density in the windings and the magnetic forces on the superconducting coils. In this process crucial constraints were

- Leave enough room in the valley between the sectors to install the RF cavities;
- Minimize the volume of the ring cyclotron to reduce the construction cost;
- Select a low gradient field in the valley area to allow the crossing of the injected beam without strong focusing/defocusing effects.

Table 6.1 presents the main parameters of the present study for the Superconducting Ring Cyclotron. The pattern of the magnetic field is shown in Fig. 6.1.

Table 6.1: Main parameters of the conceptual study of the SRC

| | | | |
|---|---|---|---|
| $E_{max}$ | 800 MeV/n | $E_{inj}$ | 50 MeV/n |
| $R_{ext}$ | 4.90 m | $R_{inj}$ | 1.8 m |
| <B> at $R_{ext.}$ | 1.88 T | <B> at $R_{inj}$ | 1.06 T |
| Bmax | <6.3 T | Pole gap | 60 mm |
| $\xi_{spiral}$ | < 12° | Flutter | 1.4 - 1.97 |
| Coil size | 17 x 27 cm$^2$ | $I_{coil}$ | 5000 A/cm$^2$ |
| Outer radius | ≤7 m | Hill width | 23° |
| Sector height | 5.6 m | N. Sectors | 8 |
| Sector weight | < 500 tons | N. Cavities | 6 |
| 4 Cavities | Single gap | 2 Cavities λ/2 | Double gap |
| Acc. Voltage | 550 - 1000 kV | Acc. Voltage | 200 - 250 kV |
| Power/cavity | 350 kW | Power/cavity | 300 kW |
| RF | 49.2 MHz | Harmonic | 6$^{th}$ |
| <ΔE/turn> | 3.6 MeV | Number of turn | 420 |
| ΔR at $R_{ext}$ | 5 mm | ΔR at $R_{inj}$ | > 10 mm |

---

[†] The cyclotron frequency is $\omega = B(q/\gamma m)$, or the revolution frequency for a particle of mass "γm" depends only on the value of the magnetic field, and the particle charge. Though B varies around the orbit through the "hills" and "valleys" of the cyclotron magnet design, the revolution time of the particle depends on the integral or average of the field over a full orbit. As γ increases, the average magnetic field seen by the particle must increase so it will maintain the same revolution time, or remain "isochronous" in the cyclotron.

The total current in the coils is (2 x 2.295·10$^6$ A·turn). The maximum field on the NbTi coils is about 6.1 T. The stored magnetic field energy for each sector is 334 MJ, about 42% greater than the stored energy (235 MJ) of the RIKEN SRC sectors [Okun2007]. The distance between two contiguous cryostats, at a radius of 1700 mm, is about 500 mm. This space should be sufficient to install either single-gap RF cavities ("PSI-like") or double-gap cavities. Unfortunately, due to the shape of the single-gap cavities and to the crowding at the center of the SRC, it is not possible to install more than 4 PSI-like cavities. There is, however, sufficient room to install the double-gap cavities in at least two other valleys. Indeed, the double-gap cavities have no protrusion into the center of the SRC as they reside just inside the space of the valleys. According to the analysis presented in section 9, the performance of the single-gap cavities is much better than that of the double-gap cavities.

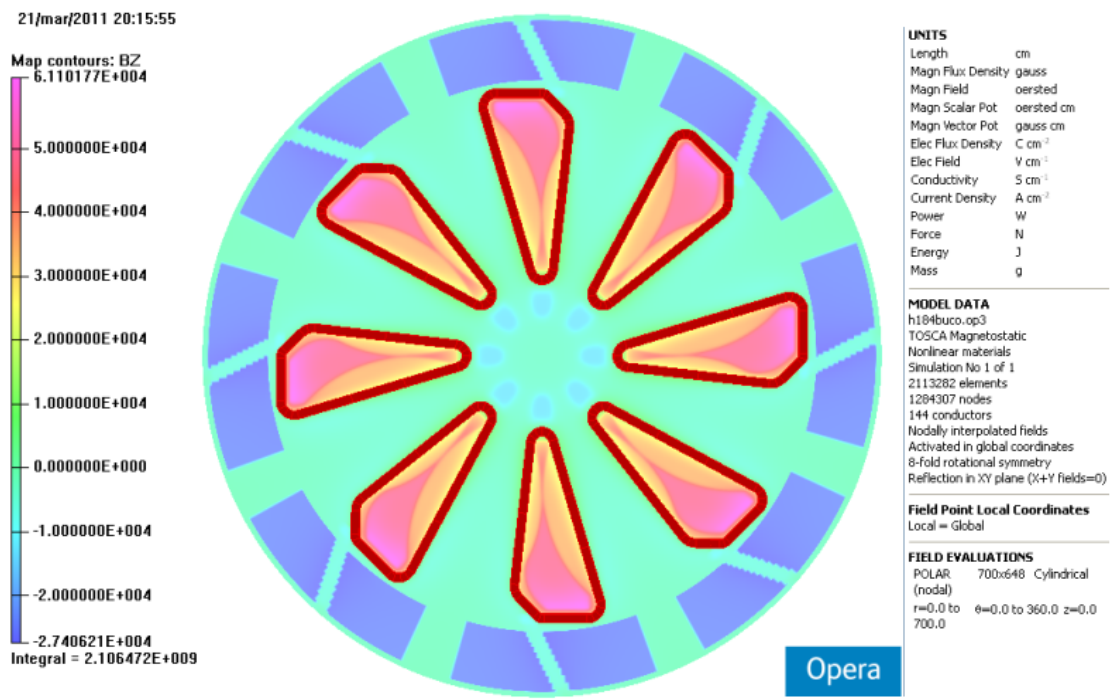

*Figure 6.1: Magnetic field map of the Superconducting Ring Cyclotron*

Despite this disadvantage, adding the two double-gap cavities along with the four single-gap cavities is very helpful. It increases the energy-gain per turn; this increase is crucial at the inner radii. In this region the accelerating voltage of the single gap cavities is lower, and it is helpful to increase the separation between the injection trajectory and the first accelerated orbit. Moreover, the additional cavities increase the average energy gain per turn by ~25%. In other words, the number of turns is reduced by 25%. Reducing the number of turns reduces the beam losses due to the interaction of the beam particles with the residual gas in the vacuum chamber.

In comparison with the design that was presented at the Lanzhou Cyclotron conference (Sept. 2010) [Cala2010], the main changes in the design of the magnetic sector of the SRC are:

1- The cross section of the superconducting coils is now smaller, 170 x 270 mm$^2$, but the current density was increased up to 50 A/mm$^2$. These changes allow more space in the valleys especially at inner radii. Now the minimum distance between contiguous cryostats, at a radius of 1700 mm, is about 500 mm. This value should suffice to install the single gap PSI-like RF cavities;
2- The coils of each sector are tilted by ±3° with respect to the median plane to reduce the magnetic field at the inner radii and increase the field at the outer radii;
3- The amount of iron and the absolute value of the magnetic field in the valley have been diminished to permit a simpler injection of the beam;
4- The iron of the hill nearest to the median plane has a strongly modulated shape to produce the required isochronous field with accuracy better of ± 1%;
5- The section of the hill, between the radii 4400 and 5000 mm is strongly spiralled to produce sufficient vertical focusing to maintain the vertical focusing frequency $v_z$ to be > 0.5;
6- The superconducting coils do not conform exactly in shape to the iron of the hill;
7- The increased angular width of the yoke compensates for a reduction of the yoke thickness.

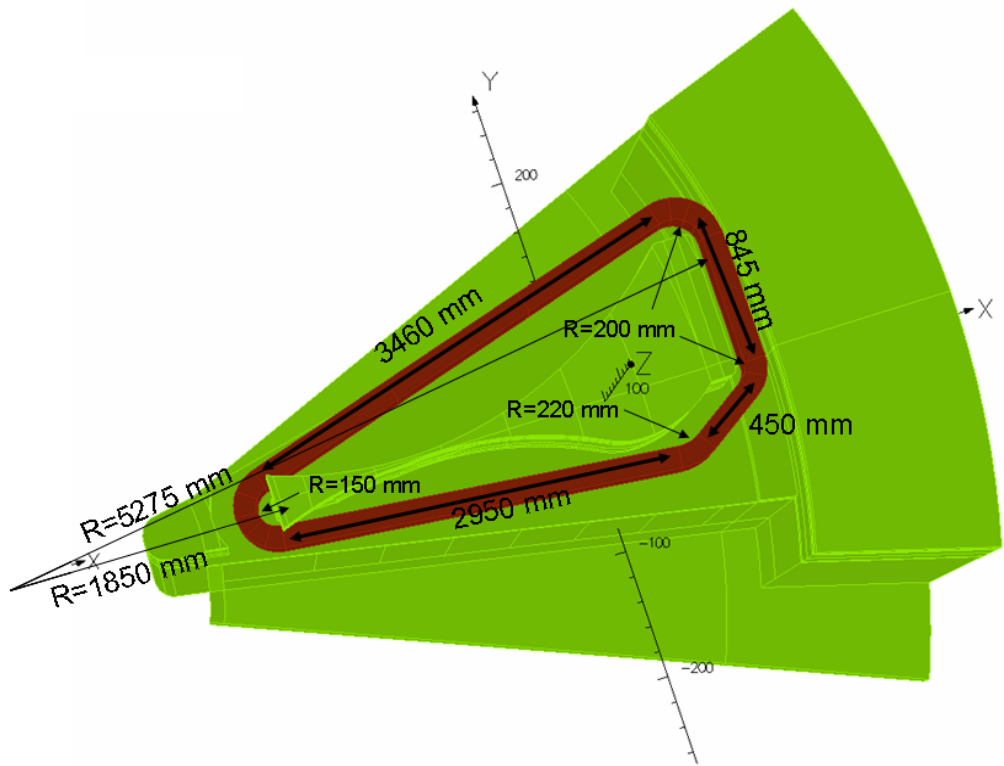

*Figure 6.2: Drawing of model H189, the pole, the length of the straight sections of the coils and the inner curvature radii of the superconducting coil are also listed*

Figures 6.2 and 6.3 show the lower half of the sector configuration called H189 with the shape of the superconducting coil as indicated. The minimum distance between the upper and lower coils, at R=5450 mm, is 150 mm, while at R=1650 mm the distance between the coils is 546 mm. This configuration leaves a minimum clearance of 35 mm for the beam at radius of 4900 mm, where the beam reaches its maximum energy before extraction. At radii smaller than 4700 mm the clearance of the vacuum chamber is everywhere > 45 mm. The 45 mm limit inside the gap between the hills is mainly due to the gap of the sector hill that is 60 mm. The upper and lower surfaces of the vacuum chamber placed between the hills of the magnetic sectors should have a thickness <7.5 mm. For safety reasons, to avoid beam halo striking the vacuum chamber, and to achieve a better conductance for the vacuum system, this clearance increases to at least 50 mm and the gap between the hills increases to 65 mm or more.

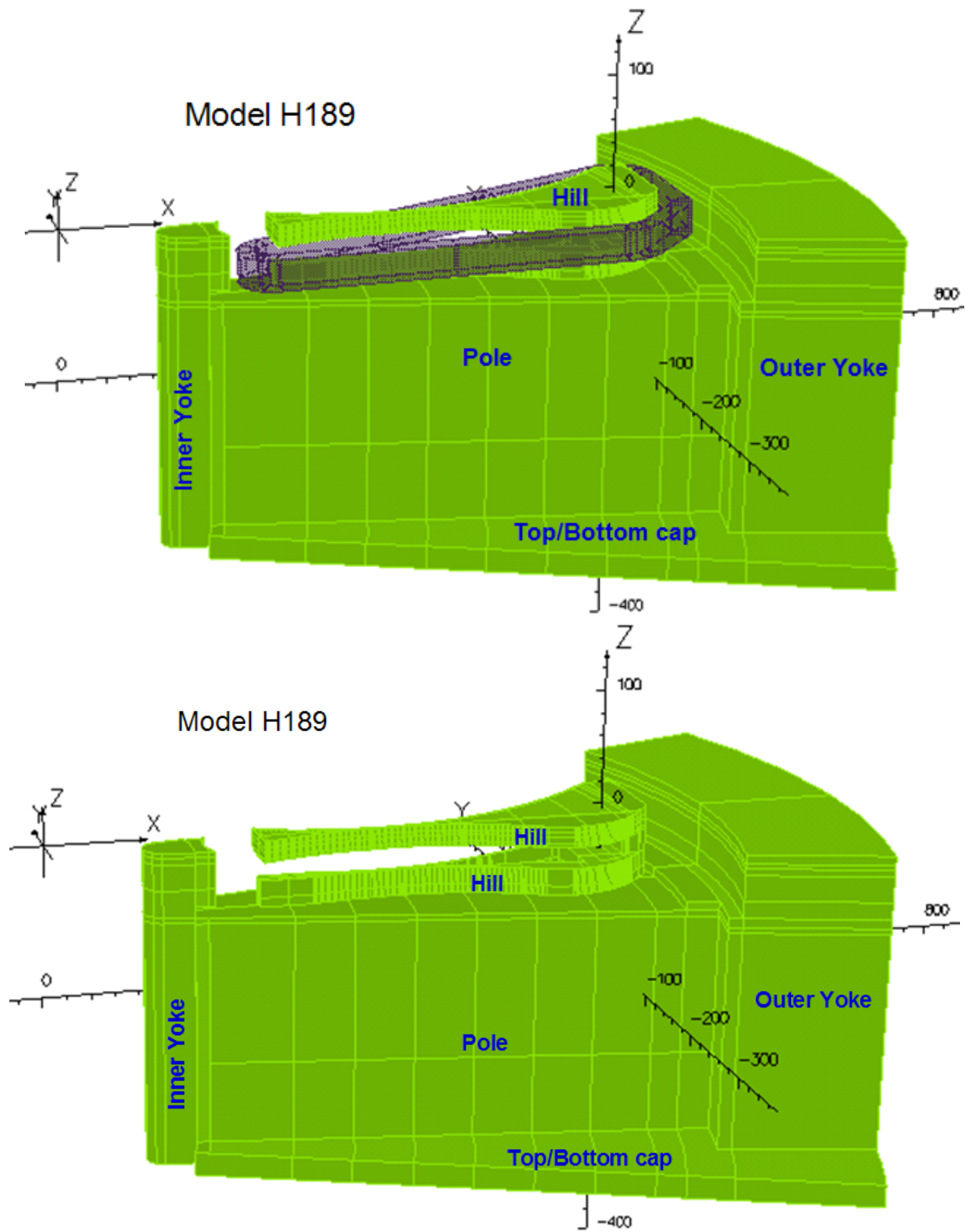

*Figure 6.3: Drawing made with the TOSCA code, showing the pole, hill, and yoke of the bottom half of one sector, with and without the superconducting coil.*

At the other end, at the inner radii, the large distance between the coils simplifies the insertion of the RF cavities and in particular of the double-gap cavities which should have a DEE width larger than 27°. The ideal value of the DEE width for a double-gap cavity working at the $6^{th}$ harmonic is 30°. In this case, the voltage gain across the double-gap is just the voltage of the DEE at that radius.

If the DEE width is 27°, the voltage gain across each gap is 98.5% of the voltage on the DEE.

The current density in each NbTi coil is 50 A/mm$^2$. Figure 6.4 shows a pair of coils. The length of each coil is divided into 9 sections, each identified by a number. In the present design iteration, the sections numbered 3, 4, 7 and 9 are assumed to be straight, but the example of the RIKEN Superconducting Ring Cyclotron suggests that in the next iteration of our design study the straight sections should be replaced with new sections having the proper bending radius to optimize the force distribution on the liquid helium (LHe) vessel.

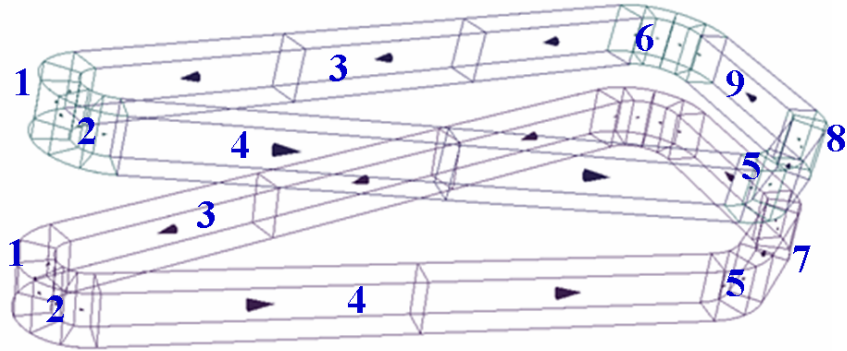

*Figure 6.4: The superconducting coil is simulated by the sum of 9 segments.*

The OPERA code evaluates the force on each coil section. The main parameters of the coils and forces which act on each section of the coil are presented in Table 6.2, which shows the components of the magnetic forces along the x,y and z axis. The reference coordinate frame of OPERA, shown in fig.6.2 and 6.3, is used. The components of the forces are along the axis x, y and z of a canonical reference frame with origin in the median plane and at the center of the cyclotron. The x-axis is directly perpendicular to segment 9 of each coil. The z-axis is perpendicular to median plane of the cyclotron.

Table 6.2: Main parameters of the superconducting coils

| Coil size | 17 x 27 cm$^2$ | Icoil | 5000 A/cm$^2$ |
|---|---|---|---|
| Max. Field | 6.3 T | Max. Force | 4.4 x 10$^7$ Newton |

| Coil Part | Length (cm) | $F_x$ (x10$^7$ N) | $F_y$ (x10$^7$ N) | $F_z$ (x10$^7$ N) |
|---|---|---|---|---|
| 1 | 32 | -95 | 69 | -23 |
| 2 | 32 | -99 | -86 | -24 |
| 3 | 340 | -350 | 1,164 | 96 |
| 4 | 290 | -127 | -1,081 | 75 |
| 5 | 21 | 33 | -104 | 34 |
| 6 | 43 | 109 | 151 | 75 |
| 7 | 49 | 145 | -157 | 76 |
| 8 | 18 | 64 | -28 | 31 |
| 9 | 88 | 390 | 0 | 177 |

The magnetic forces on the superconducting coils are large due to the high magnetic field and the large size of the coils. Sections 3 and 4 of the coils are the longest and are pushed outward by a strong magnetic force, which tries to make the coil round. To balance these forces, the wall of the LHe Vessel must be thick. The present design assumes a thickness of 4 cm for the inner wall and for the plate near the median plane. For the outer wall, and for the plate far from the median plane, a thickness of 7 cm was chosen, (see fig. 6.5).

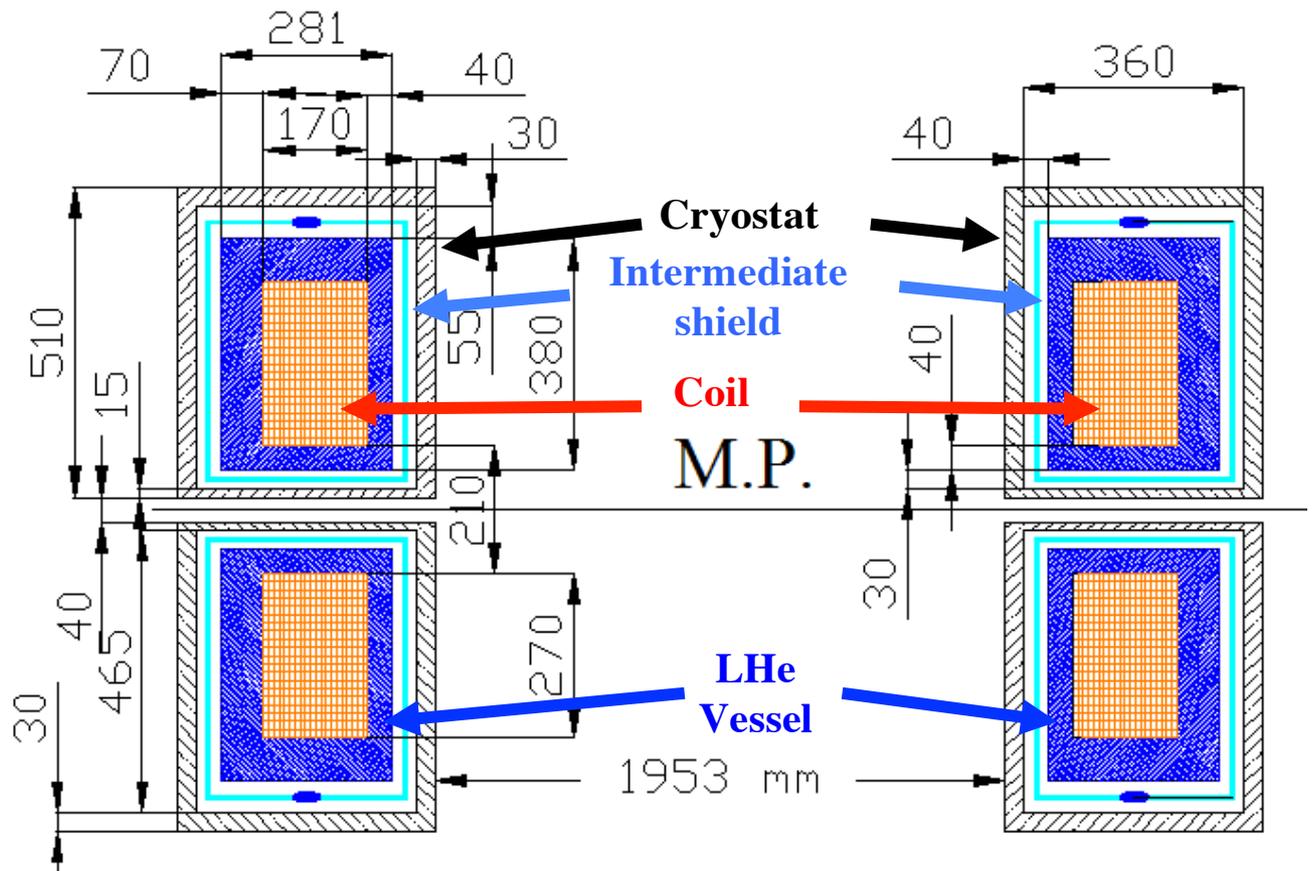

*Figure 6.5: Layout of the cryostat at radius R=4860 mm. The median plane is unobstructed to allow the crossing of the beam. The cryostat, the insulation vacuum chamber with the intermediate shield, the liquid helium vessel and the coils are shown.*

    The structure of the LHe Vessel cannot sustain the large component of the magnetic force acting on sections 3 and 4 in the y-direction. This force component requires the design to accommodate ~1000 tons of force. Fortunately, the RIKEN design group has already faced and solved this problem. Their elegant solution consists of connecting the two sections, 3 and 4 by a stainless steel plate inside the cryostat (see fig. 6.6). The forces on sections 3 and 4 have roughly the same magnitude but opposite directions; if the two sections are joined by a connecting bar, the resultant force, $F_y$, is about null. Fig. 6.6 shows the cold roughly-90 mm thick connecting plate, of AISI 316L stainless steel, that joins the LHe vessel of sections 3 and 4 of each coil. The connecting plate could start at radius 2.2 m and end at radius 4.5 m, with a full radial length of 2.3 m.

    The size of the connecting plate can be changed according to the configuration of the sector and of the coil. To allow the installation of the connecting plate inside the hill of the sector there is a gap of 190 mm from the inner radius of the hill to R = 4950 mm. It is centered at 305 mm from the median plane. The empty box of 190 mm height is enough to host the connecting plate of 90 mm of thickness and an empty area of 40 mm all around the plate for the intermediate shield and the cryostat.

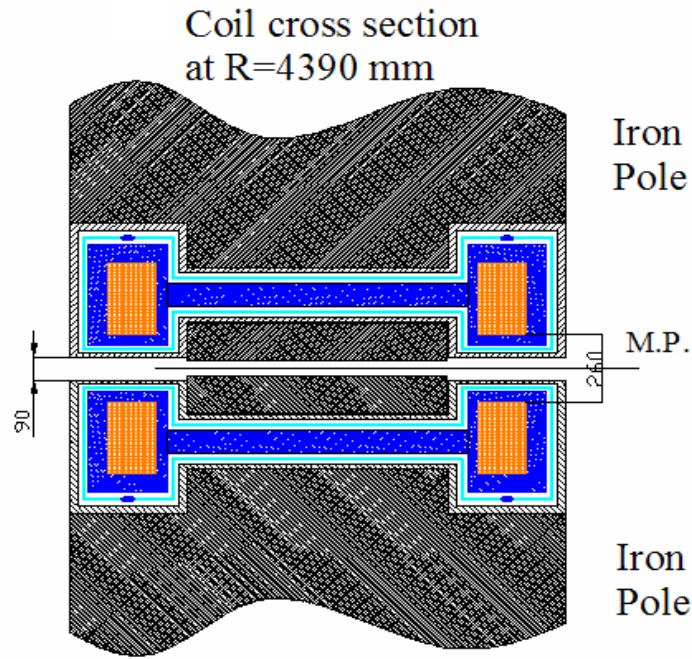

*Figure 6.6: layout of the cryostat and the inside coils across sections 3 and 4 of the coil. The LHe Vessel and the connecting plate are made of AISI 316 L stainless steel. The position of the iron pole and of the Hill iron are also shown.*

Cryogenic insulation between the 4 K connecting plate and the magnet steel requires a complex cryostat design. To assist in preserving the magnetic field, the cryostat wall that faces the pole base is made of iron. The outer parts that face the valleys and the median plane are made of stainless steel. The use of both iron and stainless steel to build a cryostat has been already demonstrated. Even if this solution introduces some technical issues, it is safe and reliable. This solution also optimizes the magnetic field configuration. Figure 6.7 shows additional views of the two ends of the cryostat. Corresponding to the coil sections 1 and 2 at inner radii and to the coil sections 6,8 and 9, at outer radii, the upper and lower part of the LHe vessels are in contact to counterbalance the huge attractive magnetic force between the upper and lower coils.

The thicknesses of the LHe vessel and of the connecting plate are chosen to maintain, throughout, stress values less than 60 MPa (60 N/mm$^2$). The region between the LHe vessel and the cryostat walls is in vacuum to preserve cryogenic insulation. The distance between the inner wall of the cryostat and the outer wall of the LHe vessel is fixed at 40 mm and is sufficient to accommodate more than 30 layers of aluminized-mylar superinsulation wrapping.

To minimize the thermal load on the LHe vessel, an intermediate shield, cooled to 77 K is installed in the gap between the cryostat and the LHe vessel. Enough room is included above and below this shield for an elliptical (35 x 15 mm) cooling tube for liquid nitrogen (see fig. 6.5). Figs. 6.5 and 6.6 show the empty region at the mid plane, between the two halves of the cryostat to house the vacuum chamber that contains the beam. Fig. 6.7 shows the cryostat vertical sections at the location of coil sections 1 and 2 at the inner most radius of the cryostat and 6, 8, and 9 at the outer radius. As these parts of the cryostat are inside and outside the beam orbits respectively, they can be in contact connecting the upper and lower parts of the LHe vessel and help to balance the attractive force between the coils. For sections 3 and 4 of the coils the attractive forces are a serious problem. These forces have to be sustained only by the cold structure of the LHe vessel. A detailed engineering evaluation of the structure will be done when the study of the sector magnet configuration is completed.

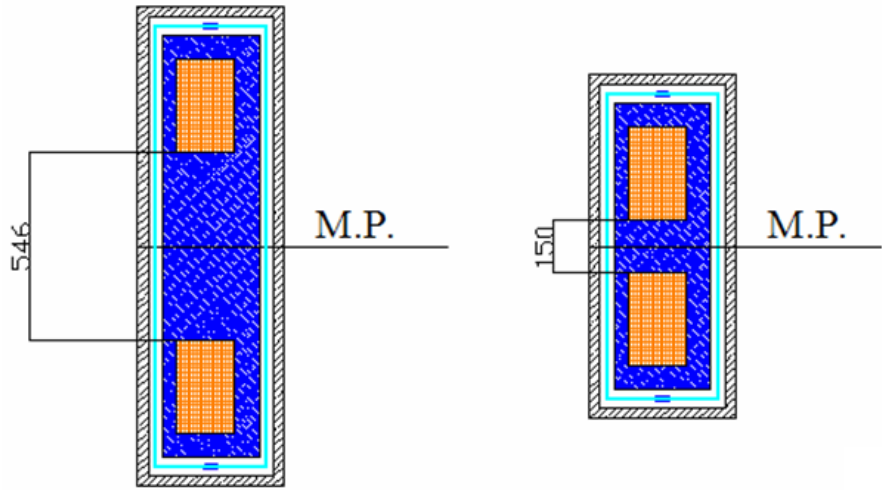

*Figure 6.7: Layout of the cryostat at the inner and outer radii. At the two ends, the upper and lower part of the cryostat are in contact. The strong attractive forces between the two coils are balanced by the central part of the LHe vessel.*

The highest magnetic field value on coil sections 1 and 2 of the previous design [Cala2010], which was about 7.5 T, is now decreased to 6.3 T as shown in fig. 6.8. The reduction arises mainly from the 3° tilt angle of the coil, which also provides some reduction in magnetic forces at the inner radii. Nevertheless the magnetic field value is still a small amount higher than the accepted maximum for the NbTi conductor, due to the small bending radius of the sections 1 and 2.

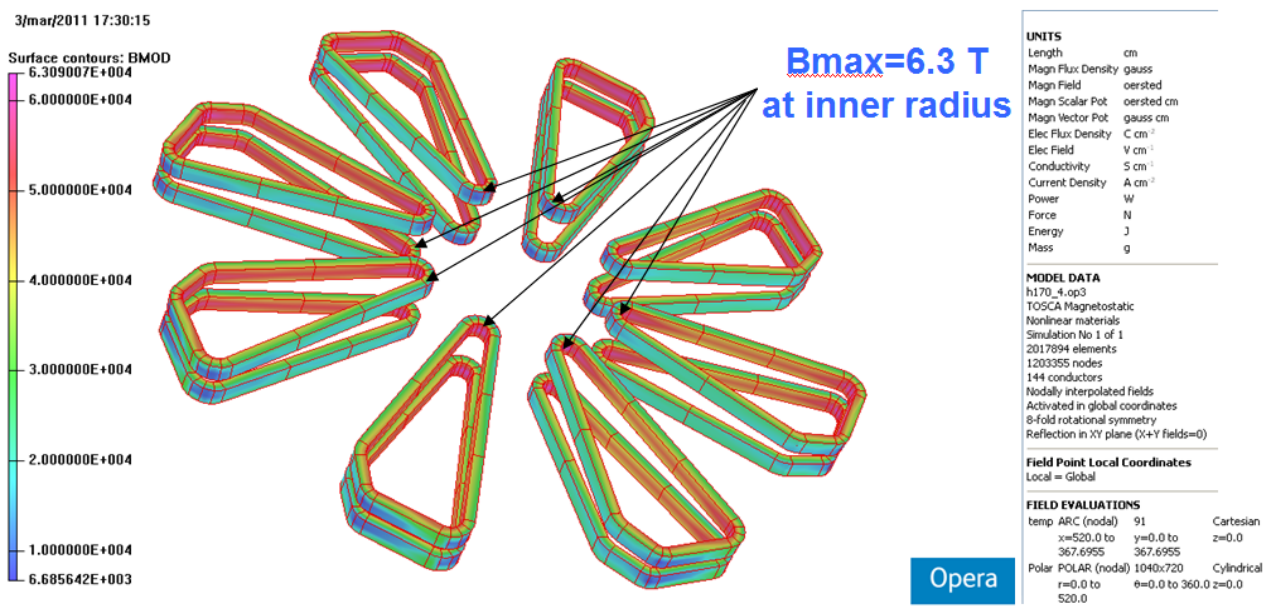

*Figure 6.8: Magnetic field on the surfaces of coils*

The total radial force in the plane of the coils is extremely high, of the order of 100 tons or more. It is directed towards the outer radius due to the absence of symmetry along this direction and due to the resultant effects of stray fields of the other seven sectors. However, this radial force can be minimized and, in principle, brought to zero, using a proper distribution of iron at the inner and outer radii. This procedure does not significantly change the main field in the acceleration area, but allows reducing the radial force that pushes each coil out.

This optimization is tedious and time consuming and needs to be redone each time any part of the pole or coil is modified. For this reason, at this stage of the design work it is not productive to spend much time in this optimization. However, the experience of the RIKEN SRC suggests that even when completed, we can still expect a residual radial shifting force in the order of 20-40 tons. In the RIKEN SRC, a pillar designed to withstand a maximum force of 90 tons counterbalances this radial shifting force. We plan to insert a similar pillar in our design.

Although the coils produce a large fraction of the magnetic field, it is the magnetic field produced by the iron in the hill region that shapes the magnetic field necessary to assure the correct isochronous field. As seen in Figs. 6.2 and 6.3, the iron of the hill has an angular width larger than 23° at starting radius of 1700 mm; it then decreases smoothly reaching the minimum value of 2° at radius R=3500 mm. Beyond this radius it expands to the radius R = 4400 mm. From R = 4400 to R=5000 mm the hill starts to take on a spiralled shape. In addition to these shape modulations easily seen in Figs. 6.2 and 6.3, other, smaller adjustments are also made at the empty aperture which divides the two parts of the hill and which holds the connecting plate and cryostat. Shape adjustments of the yoke also help in overall field configurations.

All these changes have been carefully designed to achieve both a good isochronous field and sufficient vertical focusing. Figure 6.9 shows the difference between the theoretical revolution frequency and the revolution frequency evaluated using the MSU code GENSPE [Gord1986] calculated for the magnetic field configuration H189 illustrated in Figs. 6.2 and 6.3. Data sets for all of the magnetic field configurations evaluated are available on request.

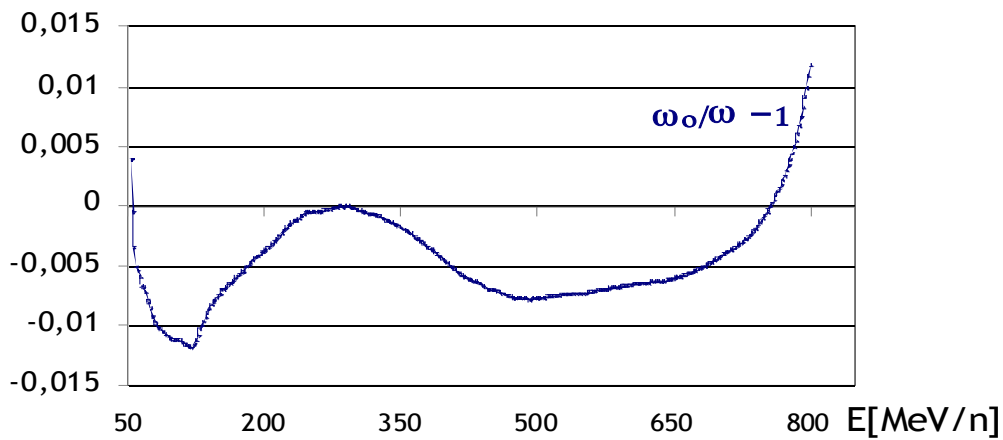

*Figure 6.9: Relative difference between the ideal revolution frequency $\omega_0$ and the revolution frequency of the particles calculated for magnetic field configuration H189.*

As seen in Fig. 6.9 the difference between the ideal isochronous field and the simulated field is less than 1% except at the extraction radius where the difference is larger, but only 1.2 %. At other radii the difference between the ideal isochronous field and the simulated field is generally less than 1% and has a minimum of -1.2% at the radius corresponding to an energy of 70 MeV/n. Negative values mean that the magnetic field must be decreased while a positive difference means that we need a higher magnet field. Obviously reducing the magnetic field is generally easier than increasing it.

The positive difference of 1.2% in Fig. 6.9 means that at the extraction radius we must increase the magnetic field by this amount. Reducing the pole gap in the outer region of the hill where the beam does not cross the pole, produced no significant change. Enlarging the angular width of the coil in the outer region, is deleterious, because it reduces the flutter of the field and the vertical focusing power.

Our strategy at this point was to increase the current in the coil. Raising the current density raises the field in the central region, where the average field due to the coil is already high, to a level that exceeded the limit of 6 T. Moreover, as the current density in the coil became too high, we tried increasing the size of the coil. This approach, however, has the drawback of reducing the free space available for the RF-cavities at the inner radii, and the high field on the coil surface still persists.

Another approach may work better. We are exploring placing an additional coil in the region from R=4500 mm and R=5200. The additional coil need be only 4 cm thick and have the same height as the main coil. In other words, the main coil would consist of two parts: the main winding, the outer part of which is wrapped around the whole hill, and a smaller, inner winding which is wrapped around only the outer part of the hill, as shown in Fig. 6.10. In this configuration, the coil size in the outer parts is 200 x 300 $mm^2$ with an area 30% higher than the original value of 17x27 $cm^2$. Thus the current density can be decreased by 20% from 50 $A/mm^2$ to 40 $A/mm^2$. Then, the total resulting ampere·turns would increase by 4.5%. The expected average magnetic field should be about 2% higher, enough to recover the 1.2%, needed according to Fig. 6.9. At the same time the coil at the inner radii has a smaller size, 16x30 $cm^2$, and a smaller current density with respect to the original value. The expected field would be 20% lower, thereby reducing the peak magnetic field on the coil.

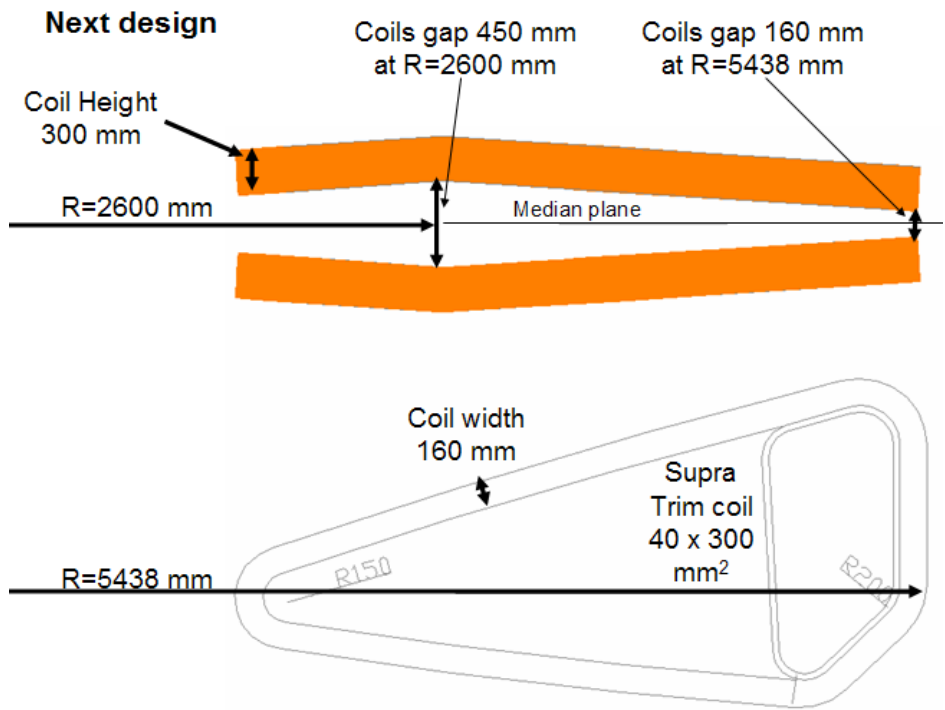

*Figure 6.10: Tentative layout of the new superconducting coil with one additional superconducting trim coil. New current density is 40 $A/mm^2$.*

To compensate for the strong reduction of the magnetic field we must bring the coil nearer to the median plane by bending the coil as shown in Fig. 6.10. The coil is shown relatively parallel at the median plane from the inner radii up to R=2600 mm. It then is tilted to achieve the separation between upper and lower coil of 160 mm at the outer radius. The minimum distance is 10 mm larger than in the configuration of Fig. 6.4 and 6.7. Moreover, the tilt angle of the coils at the outer radii will be larger than before, while at the inner radii the coil has a smaller size, 16x30 cm$^2$, and a smaller current density than originally. Moreover, the tilt angle of the coils at the outer radii will be larger than the previous ±3°; so that the clearance of the vacuum chamber in the region crossed by the beam should be everywhere larger than 50 mm. Also, the first part of the coil between the R=1500 and R=2600 mm is tilted in such a way to help improve the isochronicity of the field in this region.

We have also verified that the insertion of the additional coil at the outer radii does not adversely affect the vertical focusing properties. A further advantage is that the use of the trim coil produces a significant reduction of the radial magnetic force.

In the future, we need to increase the accuracy of the magnetic field from +/-1% to +/-0.01%. However, as the engineering design of the coil could produce a significant change in the magnetic field map, it is not useful to search for the 0.01% precision now. The present field map has sufficient accuracy to permit the evaluation of the injection and extraction dynamics, to start the engineering study of the superconducting coil, and to evaluate the mechanical challenges related to the sector such as the mechanical deformation due to the magnetic force.

**6.2 Beam dynamics**

Fig. 6.9 compares the theoretical isochronous magnetic field and the average magnetic field produced by the sector configuration H189. The difference between the ideal isochronous field and the real average field is everywhere smaller than ±1.% except at the radii corresponding to the energies 70 and 800 MeV/n, where the difference is -1.2 and +1.2% respectively. The frequencies of radial and vertical focusing, $\nu_r$ and $\nu_z$, vs. energy, evaluated for the H189 model, are shown in Fig. 6.11, Fig. 6.12 plots $\nu_z$ vs. $\nu_r$

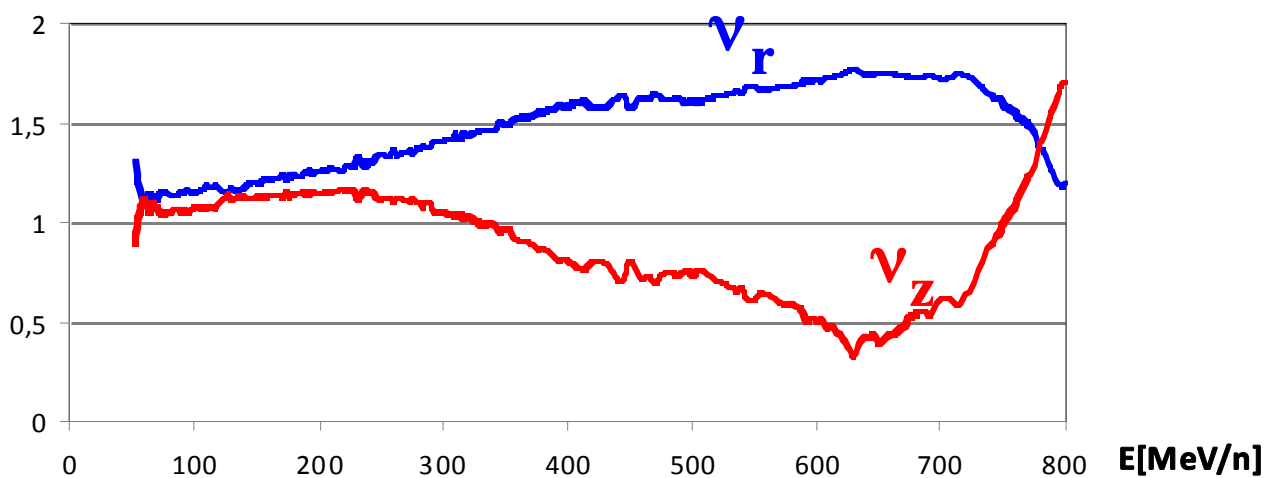

*Figure 6.11: Plot of $\nu_z$ and $\nu_r$ vs. energy E, along the acceleration path for magnetic configuration H189.*

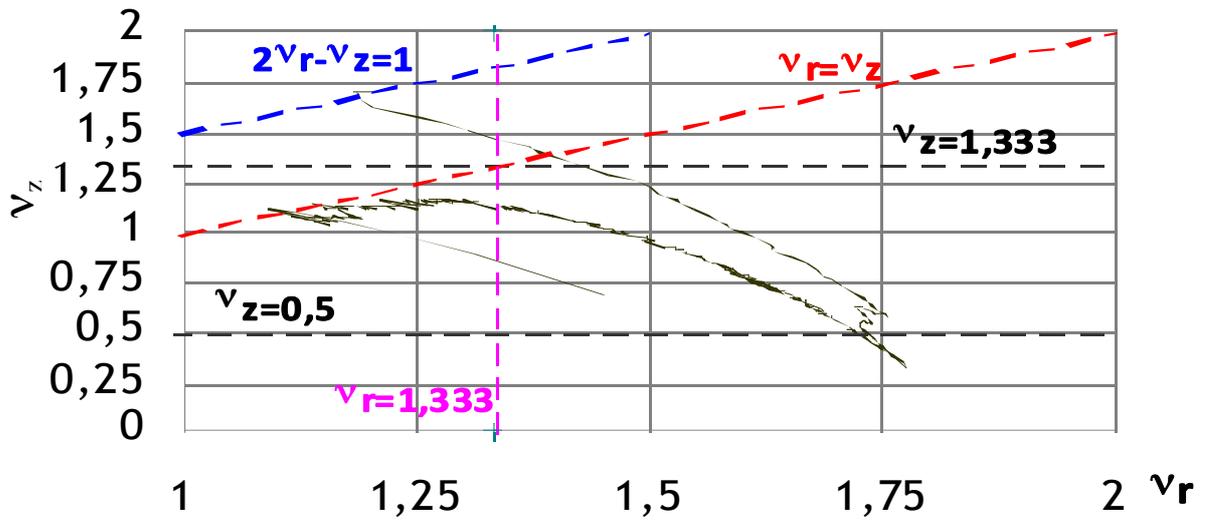

*Figure 6.12: Plot of $\nu_z$ vs. $\nu_r$ along the acceleration process. Main resonances are also plotted as dashed lines.*

The fast oscillations present in fig.6.11 are due to the grid size of the magnetic field map used to evaluate the model H189 and are not physical and not serious. Both the $\nu_r$ and the $\nu_z$ are quite regular. Moreover, the $\nu_z$ is always positive. It is a little higher than 1 at the beginning, where the space charges effects are most relevant and then decreases below 1. For a short energy interval in model H189, the values of $\nu_z$ are less than 0.5 reaching a minimum of 0.35. Calculations using the model H184 (Figs. 6.13 and 6.14) show values of $\nu_z$ to exceed 0.7. Unfortunately for model H184, the discrepancy between the ideal revolution frequency and the real revolution frequency reaches 2%. Adding the extra degrees of freedom provided by the trim coil described in the previous section, we expect to be able to meet both the goals of good isochronism and good vertical focusing. Overall, the plots of the $\nu_z$ vs. $\nu_r$, for the models H184 and H189 are very similar. It should be noted that for both Fig. 6.12 and 6.14 the dangerous resonances are crossed quite quickly.

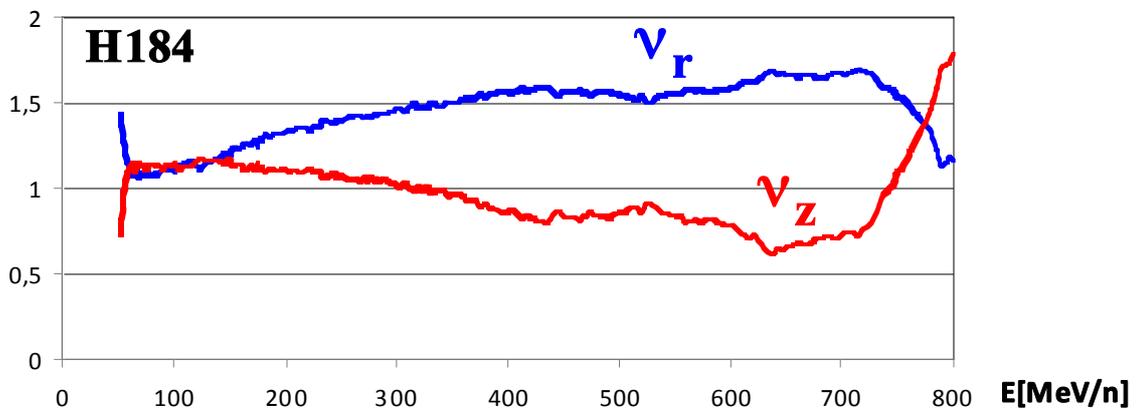

*Figure 6.13: Plot of $\nu_z$ and $\nu_r$ vs. energy E, along the acceleration path, model H184.*

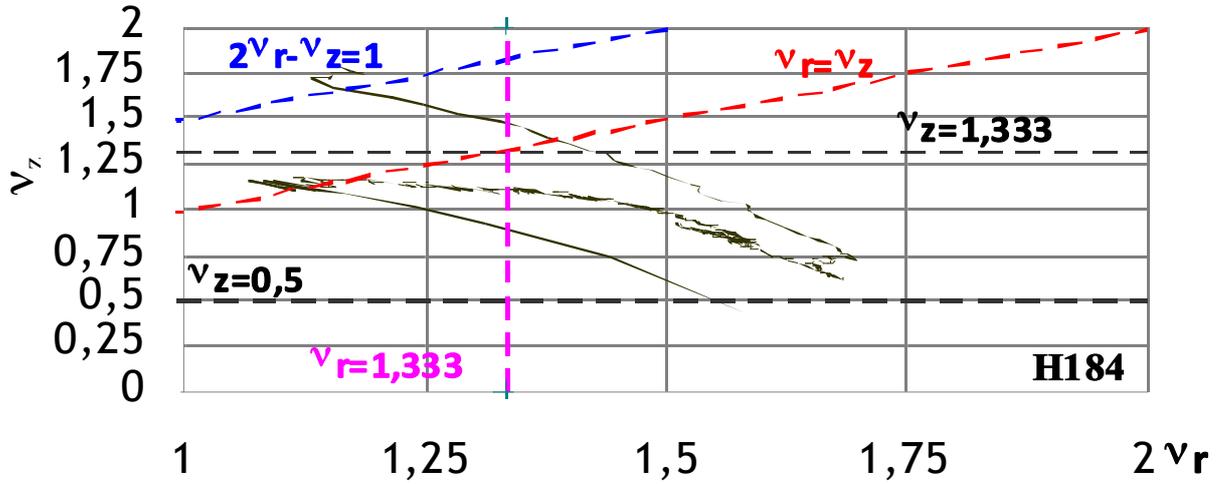

*Figure 6.14: Plot of $\nu_z$ vs. $\nu_r$ along the acceleration process. Model H184.*

## 7. RADIAL INJECTION INTO THE SRC

This section describes our simulations of different trajectories used to inject the beam into the Superconducting Ring Cyclotron. We assume the $H_2^+$ beam has energy of 50 MeV/n and a normalized emittance of 3.3 π mm.mrad. This value is about 30 times larger than the beam emittance delivered by the VIS ion source, described section 5. A conservative energy spread of ±0.2% is assumed.

The injection line must transport the beam from a point outside the yoke of the ring cyclotron to the first orbit suitable for acceleration. Fig. 7.1 shows the injection trajectory crossing the vacuum chamber of the ring cyclotron in the region of a cyclotron valley. However, even though being in a valley, the stray field is high enough to bend the trajectory. The tentative injection trajectory was simulated using OPERA. This trajectory minimizes the effect of beam blow up. However, OPERA is not the most suitable software to investigate beam injection. A better choice is a code developed at MSU to simulate extraction [Gord1986]. This code uses the main magnetic field map to describe the magnetic field of the cyclotron; along the trajectory of the reference particle the user may insert electrostatic devices and/or magnetic channels with or without a gradient field to perform fine focusing. This procedure is very efficient because one does not need to recalculate the magnetic field map for each different set up of the magnetic channel. The electrostatic and the magnetic field of the injection channel are added to the field of the magnetic map along the beam trajectory.

It is evident from Fig. 7.1 that bending the injection trajectory to match the first equilibrium orbit is not an easy task. If our injection trajectory crosses a valley containing an RF accelerating cavity, we neglect the effect of the cavity in the present study. The effect of the accelerating gap is a slight radial steering of the trajectory that is compensated by an external steering magnet.

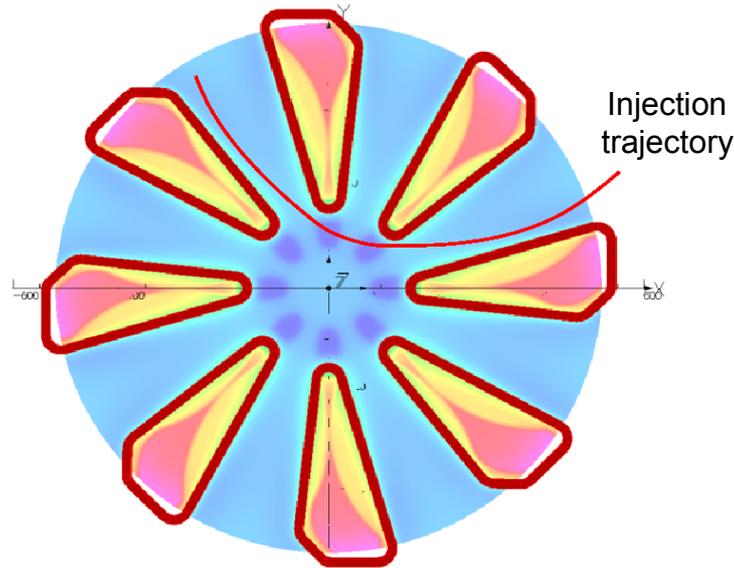

*Figure 7.1: Layout of the SRC and tentative injection trajectory evaluated in the magnetic field produced by the model H184 without any magnetic channel.*

The main constraints of the injection process are:
- The beam envelope must be small along the trajectory that crosses the cyclotron area. In our case the maximum beam envelope should be smaller than 4 cm in both the radial and vertical plane, in the region where R< 5 m ;
- The beam must match the equilibrium orbit at a reference azimuth that in our simulation is θ=0°. In the following we refer to this position as the matching point;
- The electromagnetic fields used to steer the trajectory must be realistic.

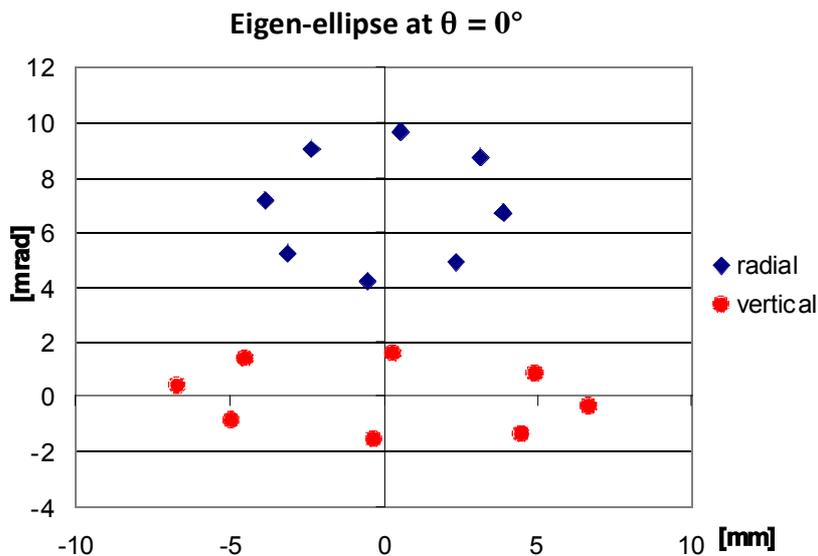

*Figure 7.2: Vertical and radial eigen-ellipse at θ=0° for the $H_2^+$ beam, with E=50 MeV/n, magnetic field map H184.*

The first step is to find the eigen-ellipse of the $H_2^+$ beam at the energy of 50 MeV/n in the main field (model H184). The normalised beam emittance area used in our simulation is 3.35 π mm.mrad; this value is about 30 time larger than the beam emittance of the VIS source. This large margin takes into account eventual non-linear effects during the injection, acceleration, and extraction from the injector cyclotron due mainly to space charge effects that increase the emittance. The eigen-ellipses are found using the GENSPE code [Gord1986] developed at MSU. The particle trajectories that describe the eigen-ellipse were then integrated to arrive at the beam envelope.

Different configurations of the magnetic channels were simulated. Here we present only two possible solutions, both of which seem reasonable. The respective trajectories are labelled A and C. The parameters of the magnetic channels (MC) and of the electrostatic deflector (ED) used to simulate the two trajectories are presented in Table 7.1.

Fig. 7.3 shows trajectory A. This trajectory was achieved using an Electrostatic Deflector (ED) placed in an empty valley just before the matching point and 5 Magnetic Channels (MC). Fig. 7.4 shows trajectory C. In Fig. 7.3 and 7.4 the electrostatic deflector is colored grey while the magnetic channels are colored red. The two trajectories are quite similar, the main difference being one more magnetic channel in A. The superconducting channel for trajectory C requires higher magnetic fields, up to 2.9 Tesla, while for trajectory A the maximum field required inside the MC is 2.7 Tesla. The ED has the same parameters for both. The inner yoke of the H184 model has a gap in the median plane of 16 cm. We left this gap both to minimize the residual $F_x$ force on the coil and also to have enough free room for the crossing of the beam. Following our simulations during the injection study, we realized that this large gap produces considerable stray field. If the gap is filled with iron, the stray field will decrease and the magnetic field to be applied at the MC should be reduced significantly.

Table 7.1: Parameters of the Electrostatic Deflector and 5 magnetic channels, placed along the injection trajectories A and C. The extraction devices ED1 and M1 are the same for both the trajectories; the other channels have some differences.

| Electrostatic Deflector | Rin (cm) | Rout (cm) | $\theta_{in}$ (deg) | $\theta_{out}$ (deg) | E (kV/cm) | Gap (mm) |
|---|---|---|---|---|---|---|
| ED1 | 183.07 | 182.29 | -8° | -38° | 50 | 10 |
| Magnetic Channel | Rin (cm) | Rout (cm) | $\theta_{in}$ (deg) | $\theta_{out}$ (deg) | B (kG) | $\Delta B/\Delta R$ (kG/cm) |
| M1 | 179.00 | 167.76 | -80° | -106° | 6 | 0.1 |
| M2 A | 160.47 | 159.14 | -118° | -124° | 12 | 0.5 |
| M3 A | 159.14 | 147.42 | -124° | -146° | 16 | 0.5 |
| M4 A | 145.69 | 140.96 | -170° | -200° | 27 | 0.5 |
| M5 A | 137.24 | 167.46 | -208° | -234° | -14 | 0.5 |
| M2 C | 160.47 | 150.69 | -118° | -146° | 12 | 0.5 |
| M3 C | 154.13 | 142.67 | -170° | -200° | 29 | 0.5 |
| M4 C | 134.19 | 159.20 | -208° | -234° | -21 | 0.5 |

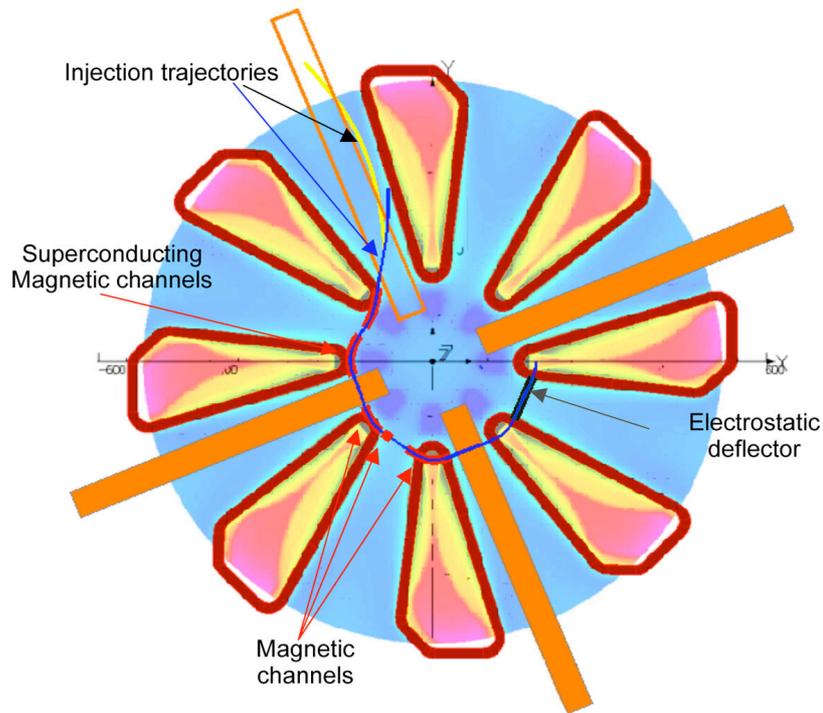

*Figure 7.3: Injection trajectory case A. The yellow line is the trajectory evaluated by OPERA. The blue trajectory was calculated with MSU's Extraction code. The positions of ED and MC are shown. The ring cyclotron layout, with the sector magnets and the 4 main RF cavities are also shown. The colors in background represent the field intensity on the media plane of the accelerator.*

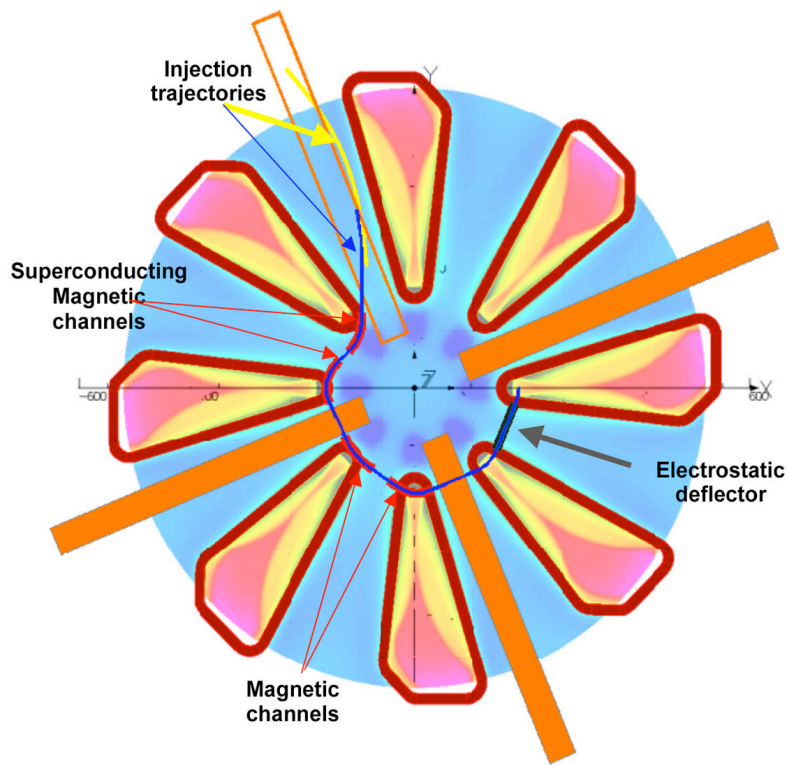

*Figure 7.4: Injection trajectory case C. The yellow line is the trajectory evaluated by the OPERA code. The blue trajectory was calculated using the MSU Extraction code. The positions of the ED and of the MC are also indicated.*

Fig. 7.5 shows the separation between the equilibrium orbit and the injection trajectories A and C. The azimuthal positions of the ED and the magnetic channels are also indicated. Note that MC 2, 3, and 4 – that require the highest magnetic field – are at a distance of about 30 cm from the first accelerated orbit. The separation between injection path and accelerating orbit is about 1 cm at the position of the ED. This distance could be increased, up to 2 cm, injecting the beam off-center by 1 cm. This first harmonic would produce a beam oscillation that increases the separation up to 2 cm at the position of the ED. However, it would be mandatory to recover from this off-center position in 2 to 4 turns. This compensation can be made by inserting a first harmonic component into the magnetic field at the proper radial and azimuth locations. The value of the voltage applied on the electrostatic deflector is 40 kV, which is very conservative.

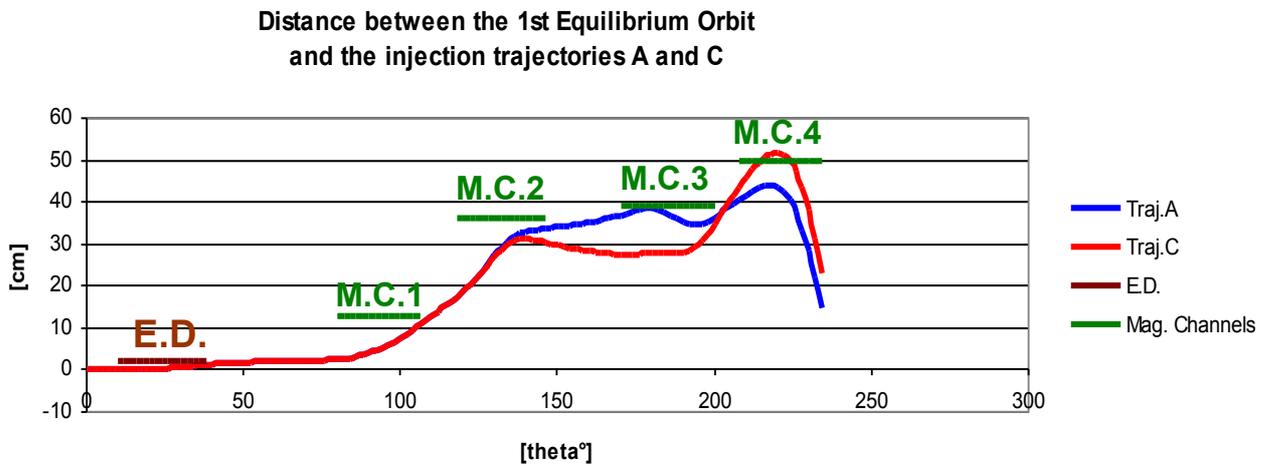

*Figure 7.5: The distances between the first equilibrium orbit and the injection trajectories. The azimuth positions of E.D. and of magnetic channels are also shown.*

Figures 7.6, 7.7 and 7.8 show the beam envelope along the trajectories C and A respectively.

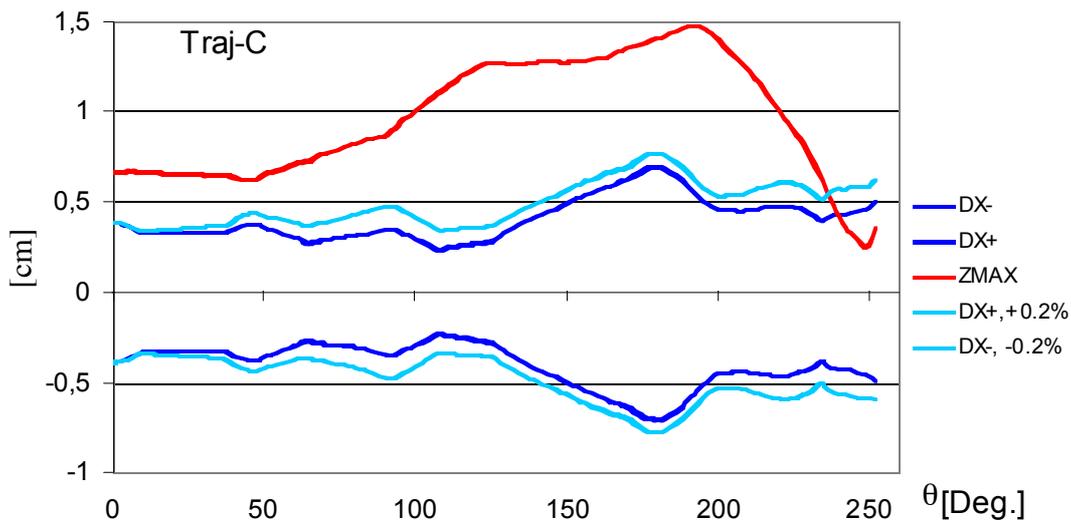

*Figure 7.6: Radial and axial beam envelope vs. θ, along the injection trajectory C. The beam envelope for a beam with energy of ±0.2% is also shown*

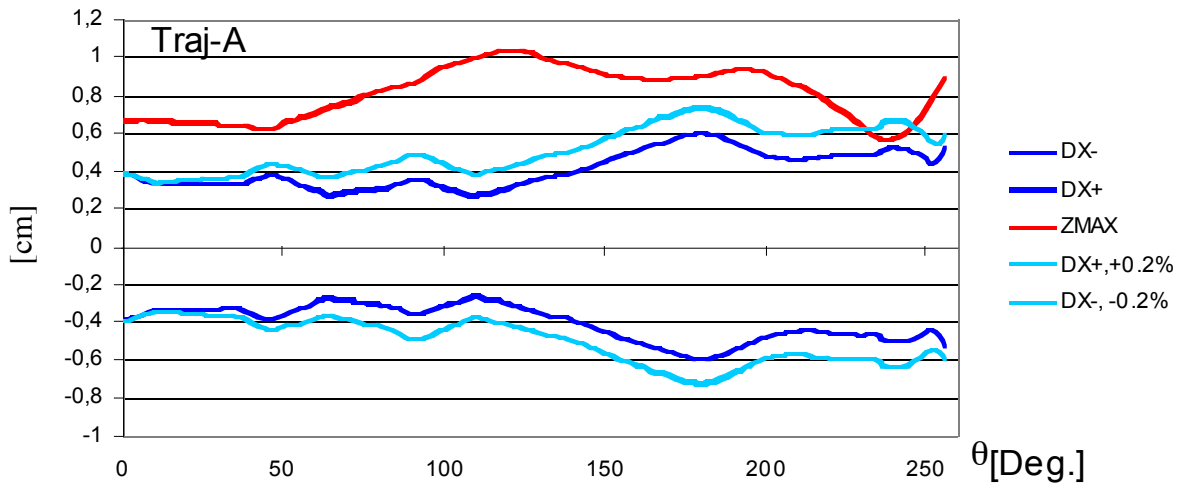

*Figure 7.7: Radial and axial beam envelope vs. θ, along the injection trajectory A. The beam envelope for a beam with energy of ±0.2% is also shown.*

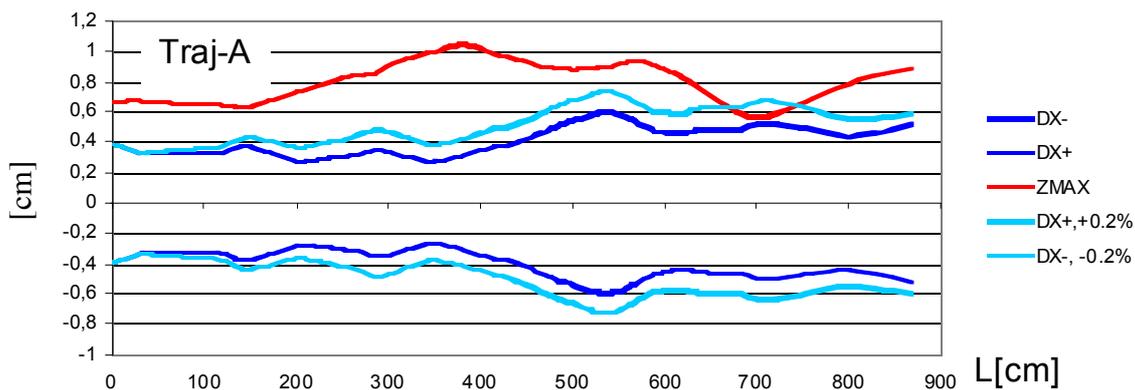

*Figure 7.8: Radial and axial beam envelope vs. L, distance along the injection trajectory A. The beam envelope for a beam with energy of ±0.2% is also shown.*

## 8. EXTRACTION BY STRIPPING

There are two standard methods to extract the beam from a cyclotron:
   1) Electrostatic deflectors produce an electric field that pushes the beam towards the outer radii until the beam reaches the region where the magnetic field falls off, and the beam escapes from the pole;
   2) The so-called "stripping method" used in the latest commercial cyclotrons that accelerate negative ions.

For the stripping method to work, the beam being accelerated must be an ionic species including one or more electrons. A thin stripper foil placed in the beam orbit inside the cyclotron will remove some, and hopefully all, of the electrons from the ions. The ion passing through the stripper abruptly changes its charge state and consequently the curvature of its orbit in the magnetic fields. The sudden change in the radius of curvature allows bringing the beam out from the cyclotron, if the stripper is properly placed and if the beam particles have enough energy. The advantage of stripping extraction is that the extraction efficiency is naturally 100%, even if the beam orbits at the extraction radius are not well separated. Therefore, the energy gain per turn need not be high as with electrostatic deflection. The disadvantage is that the energy spread of the extracted beam is greater.

Commercial low-energy cyclotrons accelerate H$^-$ ions, which when stripped bend in the opposite direction from the equilibrium orbit, and readily leave the machine. But the stripper method can also be used to extract molecular H$_2^+$, made up of two protons bound by one electron. The free protons emerging from the stripper spiral in rather than out, but as will be seen can still be easily extracted from the cyclotron. (Section 10 addresses why H$^-$ cannot be used in our accelerator.)

Fig. 8.1 shows (in green) the trajectory of the 800 MeV proton beam extracted from the DAEδALUS-SRC. This extraction trajectory was evaluated from field map of model H184 using the EXTRACTION code developed at MSU. The beam envelope vs. angular position along the proton extraction trajectory is shown in Fig. 8.2. Fig. 8.3 displays the beam envelope vs. radius along the beam trajectory. Both Fig. 8.2 and 8.3 show the radial and axial beam envelope for the 800 MeV proton beam without energy spread and with an energy shift of ±0.75%. The axial beam envelope does not change for small energy change. The maximum radial beam envelope due to the ±0.75% of energy shift is about ±6 cm.

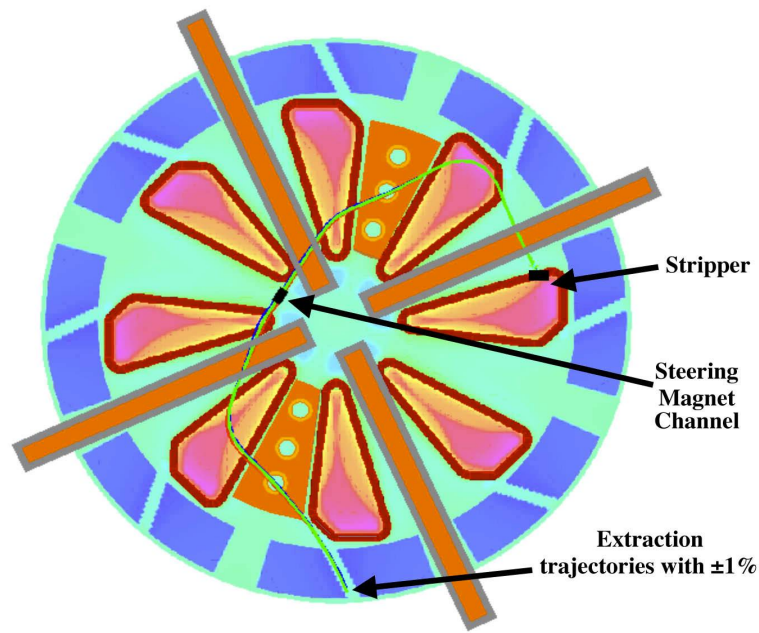

*Figure 8.1: Layout of the DAEδALUS-SRC, model H184. The extraction trajectory for the proton beam with energy of 800 MeV is shown in green. The blue and red lines are the trajectories for protons with energy ±1% from nominal, are seen faintly superposed on the primary orbit. The stripper at angular position of 15° and the steering magnet channel are also shown. For simplicity, only one extraction trajectory is shown.*

At the exit of the cyclotron the radial width of the beam envelope is less than 3 cm. The value of ±0.75% of energy spread is an upper limit; a more realistic value is ±0.26%. To obtain a rough evaluation of the energy spread we assume that the beam bunch circulating along the outermost orbits has a radial size Δx=15 mm. This value is 1.5 times larger than the beam size of the PSI ring cyclotron. Consistent with the equilibrium orbit evaluated by GENSPE using the magnetic field map

H184, and assuming an energy gain per turn of 2 MeV/n, we find a radial separation at the stripper position of about ΔR=7.3 mm. Therefore, the beam particles of the bunch will cross the stripper after different numbers of turns. Consequently the extracted beam will have an energy spread given by (ΔX/ΔR)*ΔE/E = (15/7.3)*(2/800) = 0.514% (Full energy spread) or ΔE/E=+/- 0.26 %

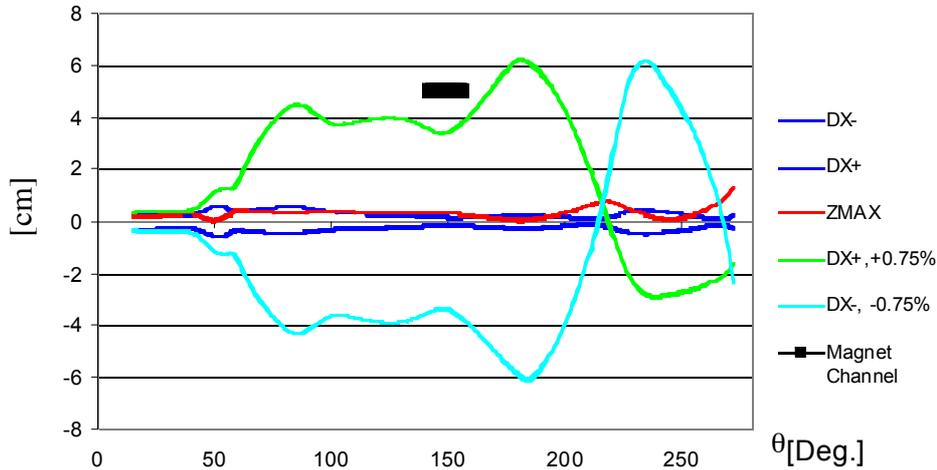

*Figure 8.2: The radial (blue lines) and axial (red line) envelope for the zero energy spread vs. angular position for the proton beam along the extraction trajectory are shown. The green line and the cyan line are the radial beam envelope for the two cases with energy spread of ±0.75%. The angular position of the steering magnet channel (black line) is also shown.*

The radial and axial beam envelopes are obtained by integrating the trajectories of eight particles which stay on the boundary of the beam eigenellipse which strikes the stripper foil placed at R=488.3 cm and at an azimuth angle of 15°. The shapes of the beam envelope in the radial and axial plane were obtained from the eigenellipse evaluated by GENSPE at the proper angle and radius of the stripper foil. The nominal energy of the beam is 800 MeV/n. The normalized axial and radial emittances are assumed to be 3.35 π mm.mrad, about 33 times larger than the ion source emittance.

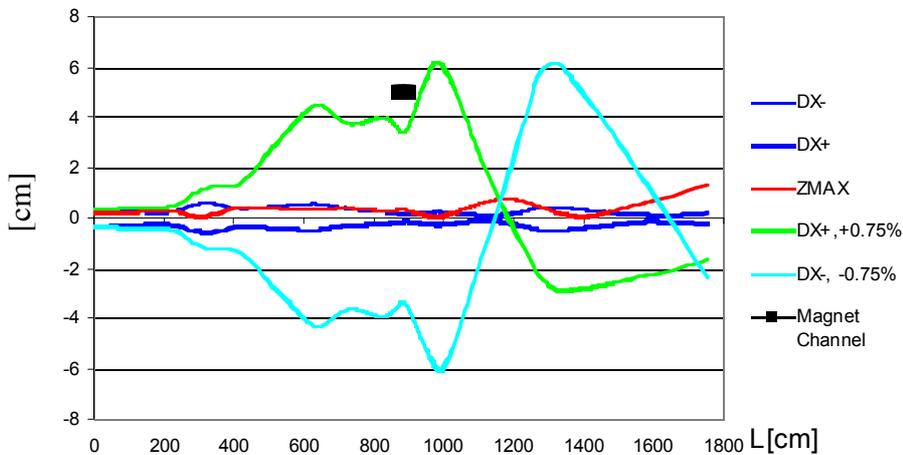

*Figure 8.3: The radial (blue lines) and axial (red line) envelope for the proton beam with zero energy spread vs. distance from the stripper position along the extraction trajectory are shown. The green line and the cyan line are the radial beam envelope for the two cases with energy spread of ±0.75%. The angular position of the steering magnet channel (black line) is also shown.*

The beam trajectory has a minimum distance from the center of 145 cm and for a narrow angular range of about 50°, the width of one sector, the trajectory has a distance from the center less than 160 cm. Therefore, this part of the trajectory remains inside the central region of the SRC. We can install a magnetic channel to steer the trajectory and also to provide axial focusing to maintain the axial beam envelope <3 cm along the entire trajectory inside the vacuum chamber of the SRC. The angular extension of the magnetic channel is from 142° to 156°, just inside an empty valley; the bias field of the MC is 0.2 T and its radial gradient is 0.2 T/cm.

The azimuth position was chosen just to place the stripper foil in the region where the magnetic field is in the range 0.2 to 0.4 Tesla. The bending radius of the stripped electron will be ~8 to 9 mm; hence these electrons can easily be stopped on a copper shield placed some millimeters after the stripper foil and 5 to 6 mm towards the outside.

Based on extensive experience with such foils, a pyrolytic graphite, highly oriented, foil will be used, with a thickness of about 2 mg/cm$^2$. This foil thickness is anticipated to strip all ions into protons. Unstripped ions will lose so little energy (~5 keV) that they will pass through the foil on a subsequent turn. Partially stripped ions (producing a proton and a neutral hydrogen) will be a very small fraction, and the neutral hydrogen atoms will strike the outer yoke in a well-defined place, this will need to receive further study.

**8.1 Stripper Mean Lifetime**

To evaluate the mean life of stripper foil we consider the experience with stripper foils used to extract H$^-$ from commercial cyclotrons and from the TRIUMF cyclotron.

According to the TRIUMF experience [Byli2010] a mean life greater than 180 mA·hour has been measured with the updated stripper foil which is 2 mg/cm$^2$ thick pyrolytic graphite (highly oriented). Generally, a "C"-shaped frame supports the stripper foil. TRIUMF now uses an inverted "L"-shaped tantalum frame. The frame supports the stripper on two sides only to avoid mechanical stresses due to erosion by the particle beam and the stripped electrons. The stripper is held in position by two pairs of screws. A further trick, to avoid mechanical stresses in the stripper foil, is the insertion of a thin wrinkled copper sheet to hold the graphite stripper in contact with the tantalum frame.

Extrapolating the TRIUMF experience to the DAEδALUS SRC, we estimate an average mean life at least of 90 hours for an average current of 2 mA. This mean life could be much longer because there are important differences between the stripping process for H$^-$ and H$_2^+$. For the H$^-$ case the two electrons must be both removed to extract one proton, and the foil has to be thick enough to guarantee stripping 100% of all the particles that cross the stripper and to minimise the losses from particle neutralisation. In contrast for H$_2^+$ if a molecule is not stripped at first crossing through the stripper, it runs for an additional turn inside the cyclotron and hits once again the stripper until it is stripped. Thus for H$_2^+$ it is possible to use a stripper with thickness less than 2 mg/cm$^2$ used for H$^-$ and then a longer mean life could be expected. Moreover, each time that a H$^-$ particle crosses the stripper 2 electrons are lost and strike the stripper foil while when the H$_2^+$ is stripped only one electron strikes the stripper foil. Hence the power due to the electrons removed by the stripper is just 50% for the H$_2^+$ vs. H$^-$ assuming the same beam current. From this simple consideration we might expect a mean life up to two times longer than what the TRIUMF experience might suggest.

A further advantage is the suppression of the electrons that strike the stripper. The magnetic field to accelerate H$_2^+$ has reversed polarity compared to the field of H$^-$, so the electrons stripped in the case of H$^-$ are bent towards the centre of the machine and hit the stripper foil after spiralling in the magnetic field, while for H$_2^+$ the electrons are bent towards the outer radius. So if the orbit radius of the stripped electrons is larger than 4 to 5 mm, an electron catcher could be installed to remove these electrons and strongly reduce the stripper damage. The extraction simulation shown in the previous section assumes that the stripper foil is installed in a region where the magnetic field is ~0.2 T. Hence, the bending radius of the electrons removed from the $H_2^+$ at the energy of 800 MeV/n is about 9 mm.

The measured mean life of the stripper foils in commercial cyclotrons at 30 MeV is driven by the electron damage of the stripper foils. If we can stop the stripped electrons, the mean life of the foil should be much longer than that measured in the small commercial cyclotron. Moreover, at higher energy the energy lost by a beam particle crossing the foil is lower than at low energy.

Recall that the energy lost by a proton beam with energy of 800 MeV in a stripper foil 2 mg/cm$^2$ thick is just 4 keV, which for a current of 4 mA translates to a power loss of about 16 W. Note, the electrons stripped from the ions carry an energy of 1.6 kW, but with proper design this energy will be deposited in a cooled catcher placed next to the foil.

The PSI cyclotron produces a pion beam by using a rotating graphite target with thickness of 40 mm and power deposition of 20 kW/mA (beam energy 580 MeV @ 2mA). The mean life of the target is exceeds 35 A-hr [Heid2002]. The PSI pion target is not directly cooled; the target is cooled only by thermal irradiation of a cooled, surrounding surface. Although there is a great difference between PSI's rotating target with a diameter of 450 mm and a fixed small stripper foil, some lessons can be drawn. The irradiated area of the rotating target of PSI is the circumference multiplied by the size of the vertical beam spot, which is 6 mm, yielding a surface area of ~8500 mm$^2$. Hence the beam power density deposited by the beam on the rotating target is of 4.7 W/mm$^2$. On the stripper foil for the DAEδALUS-SRC the beam power losses due to the proton beam interaction is 28 W and the surface of the beam is about 16 mm$^2$, so the density 1.75 W/mm$^2$, a factor 3 times smaller. From this consideration if we are can cool the surface of the vacuum chamber surrounding the foil and if a copper catcher removes all the electrons, we could expect a very long mean life for the stripper foil.

## 9. RF CAVITIES

The experience of PSI demonstrates the importance of maximizing the energy gain per turn to minimize the energy spread of the beam due to the space charge effects and consequently to minimize both the longitudinal and radial size of the beam. The radial size of the beam is crucial for a cyclotron that extracts the beam using an electrostatic deflector. Although that consideration is not relevant if the stripper method is used, using highest energy gain per turn also minimizes the length of the trajectory inside the machine. Minimizing the path length minimizes the beam losses due to the interaction of the $H_2^+$ beam with residual gas, simplifies the crossing of dangerous resonances, and lowers the required accuracy of the isochronous magnetic field.

Unfortunately it is not practical to install more than four single-gap RF cavities in an 8-sector cyclotron due to 1) the mechanical interference of the cavities in the central region and 2) the space needed for the injection line. Hence even though double-gap cavities are less efficient than single-gap cavities, we propose to add two double-gap cavities to increase the energy gain per turn by an additional 0.8 MeV per turn. The two double-gap RF cavities allow us to increase the maximum beam power by about 20%, without increasing beam losses. For these reasons the RF system of the DAEδALUS SRC should consist of four RF cavities similar to the PSI pill box cavities, which are able to produce an energy gain of ~4 MeV per turn, plus two double-gap RF cavities.

Double-gap cavities are suitable for applications in which a special radial voltage profile (along the acceleration gaps) is desired. The RF voltage must not exceed 400 kV per gap, limiting the acceleration performance and requiring a complex cooling system. The double-gap cavity option was modelled using the multi stem approach [Magg2006].

Single-gap cavities can reach higher accelerating voltages (with sinusoidal distribution) and are less sensitive to higher order modes. The cooling system braised onto the outside of the cavity is simple to fabricate with no risk of water leaking into the vacuum (high reliability)[Sigg2000]. Engineering complexity and bigger dimensions are the price one has to pay.

Both options have been analysed, and the combination of 4 single-gap and 2 double-gap cavities has been chosen as the most appropriate solution for the DAEδALUS SRC. Despite the double-gap cavities having a lower voltage than the single-gap voltage, their geometry allows easier interfacing with the magnetic injection channel in the central region. Moreover they allow the insertion of a cryopump adjacent to the median plane – an advantage with respect to increasing the performance of the vacuum system.

The single-gap cavities are pillbox resonators, practically symmetrical with respect to the median plane, and reflect the main mechanical characteristics of the PSI 590MeV cyclotron [Fitz2004]. The double-gap cavities are multi-stem $\lambda/2$ resonators symmetrical with respect to the median plane, characterised by spiralled electrodes (Dees), similar to the ones used in the superconducting compact cyclotron cavities.

**9.1 RF Specifications: RF Voltage and Power Dissipation**

The specifications of the RF cavities are presented in Table 9.1 The effective voltage required of the DAEδALUS SRC cavities is 1 MV in the acceleration gap for the single-gap design and 300 kV per gap for the double-gap cavities. Across a double-gap cavity, the beam will be accelerated twice for each cavity crossing. Each cavity must be designed to dissipate at least 500 kW (real value). Each cavity will be fed by an independent RF power amplifier. A process of performance optimization has led to modification of some geometrical parameters of the final cavity.

Table 9.1: SRC RF Cavity Specifications

| Resonance frequency | ~51 MHz |
|---|---|
| Accelerating Voltage per Single gap Cavity | ~1 MV |
| Accelerating Voltage per Double gap Cavity | ~600 kV |
| RF Thermal Power Dissipation per Cavity | < 500 kW |
| Harmonic number | 6 |
| Material | OFHC Copper |

At these very high excitation voltages management of RF leakage may be an important design consideration. Based on PSI experience [Adel2011] relating to the importance of this issue, and using computational techniques developed there [Adel2005][Bi__2011], 3D electromagnetic modelling codes will assess heat loading and RF effects on beam and components due to cavity leaking and geometric configurations.

## 9.2 The cavity geometry

*Single Gap Cavity (SGC)*

The drawings of the cavity were made starting from the final drawing of the magnet with AUTODCAD 2010 [Auto2011]. Fig. 9.1 shows a 2D top view of the SRC with effective space for the RF cavity. Due to the presence of the cryostat, a single-gap cavity with a straight section is not feasible. The radial width is modulated consistent with the effective space available.

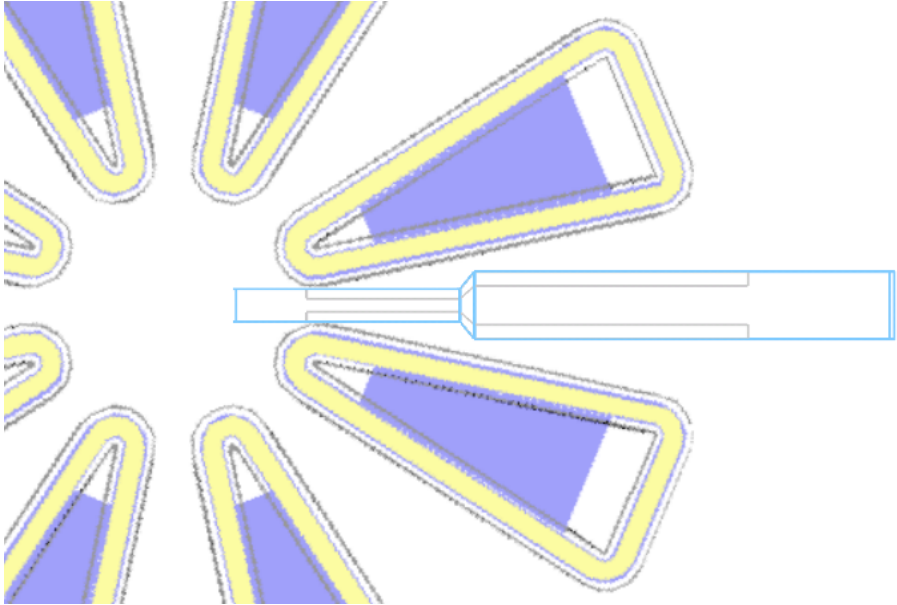

*Figure 9.1: Top 2D view of the SRC, showing space available for an RF cavity*

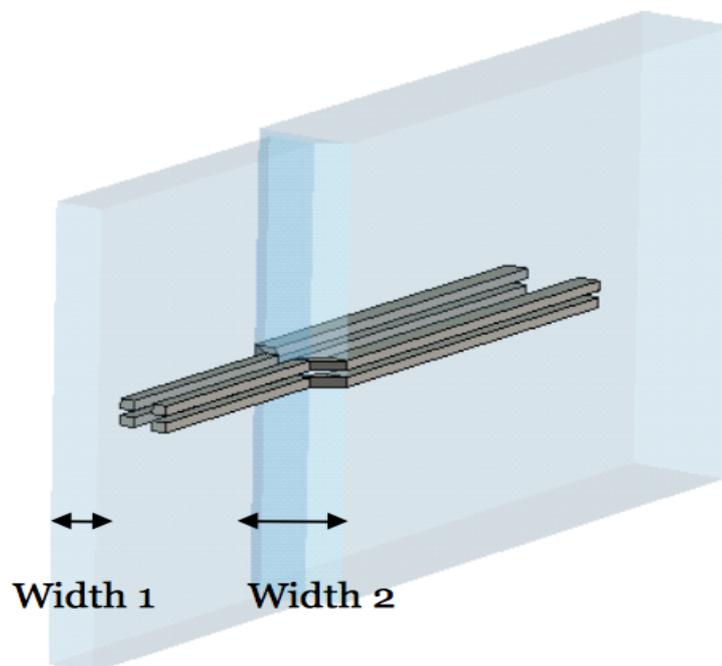

*Figure 9.2: Single Gap Cavity with modulated section.*

A single-gap cavity with two sections along the radius was designed and optimized to fulfil the performance requirements. The cavity wall consists of an 8 mm copper sheet on which cooling channels are directly TIG brazed. The large number of channels provides efficient water-cooling with very small thermal gradients. The total height of the pillbox cavity is 3000mm. The cavity radial extension is from radius 900 mm to radius 7700 mm.

The width of the cavity at different radii is:

- 340 mm from inner radius up to radius 3200 mm,
- 700 mm (as for PSI cavity) from radius 3200 mm up to radius 7700 mm.

The acceleration gap consists of two copper bars with 44 mm of clearance. Due to the radial modulation of the width, the acceleration gap is also modulated:

- 140 mm from radius 1900 mm up to radius 3200 mm,
- 400 mm (as for PSI cavity) from radius 3200 mm up to radius 4900 mm.

Fig. 9.3 shows a 3D view of the cavity, while Table 9.2 presents the main parameters of the single-gap cavity.

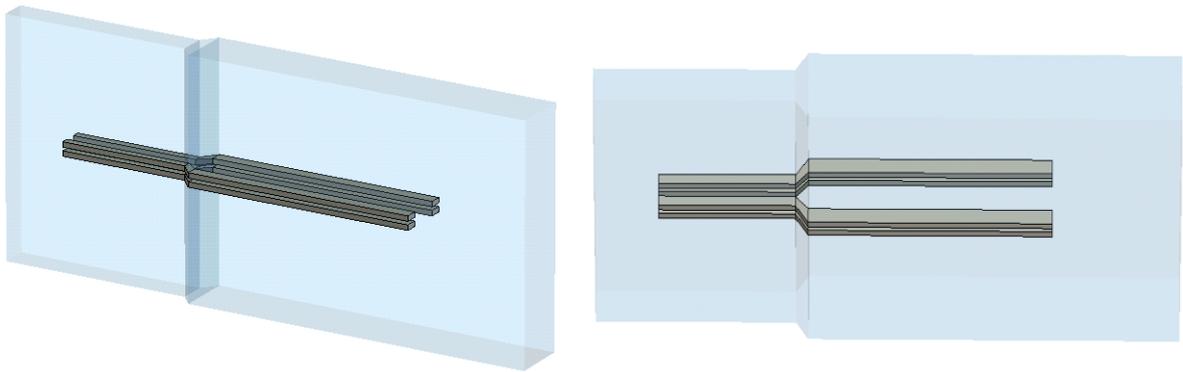

*Figure 9.3: 3D view of the SRC Single Gap Cavity*

Table 9.2: SRC SGC parameters

| Total height | 3000 mm |
|---|---|
| Radial width (radius < 3200 mm) | 340 mm |
| Radial width (radius > 3200 mm) | 700 mm |
| Acceleration gap (radius < 3200 mm) | 140 mm |
| Acceleration gap (radius > 3200 mm) | 400 mm |

*Double Gap Cavity (DGC)*

A Double Gap Cavity with three sections along the radius was designed and optimized to fulfil the performance requirements. The cavity is a resonator with three stems that are connected to the liner and the Dee. The shape of the Dee and the position of the three stems were optimized to fit the space requirements and to satisfy the RF specifications. The simulations here presented were done for a previous design of the RF cavities, for which the injection radius was 1.4 m and the extraction radius was 4.2 m. The present design calls for an injection radius of 1.8 m and an extraction radius of 4.9. Despite these values being quite different, the radial range is changed only 10%, from 2.8 m up to 3.1 m. Hence, the sizes of the DGC do not change significantly.

The total height of the Double Gap Cavity is 800mm. The cavity radial extension is about 3000 mm. The radial width of the cavity ranges between 13° in the inner region to 16° in the extraction region; thus, the acceleration gap (the distance between the Dee surface and the liner surface) ranges between 20 mm and 50 mm. The Dee gap (the vertical aperture or distance between the two faces of the Dee) is set to 30 mm, as the thickness of the Dee. The three stems are located at the proper distance to achieve the desired acceleration voltage distribution. The radius of the three stems are 80 mm, 200 mm, and 300 mm for the first, second and third respectively.

A system of flanges was designed in order to accommodate the mechanical connection of the stems to the liner and the Dee. Fig. 9.4 shows a 3D view of the cavity, while Table 9.3 lists the main parameters of the double-gap cavity.

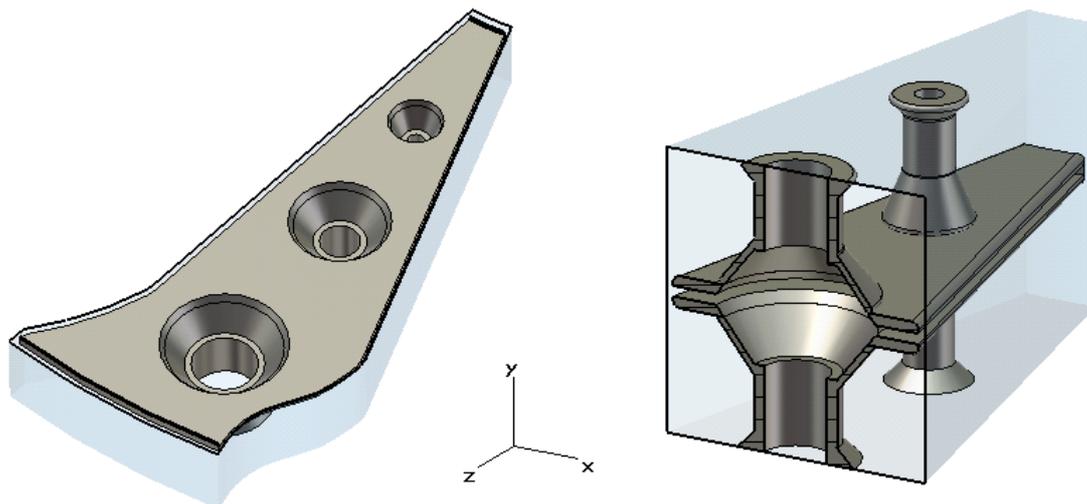

*Figure 9.4: 3D section views of the SRC Double Gap Cavity.*

Table 9.3: SRC DGC parameters

| Cavity height | 800 mm |
|---|---|
| Radial width (radius < 3200 mm) | 340 mm |
| Cavity Angular Extension (Width) | ≈ 13°-16° |
| Cavity Radial Extension (Length) | ≈ 3'000 mm |
| Dee Gap | 30 mm |
| Dee Thickness | 30 mm |
| Acceleration Gap (Dee surface - liner surface) | 20-50 mm |
| Stems Inner Diameter | 80/200/300 mm |

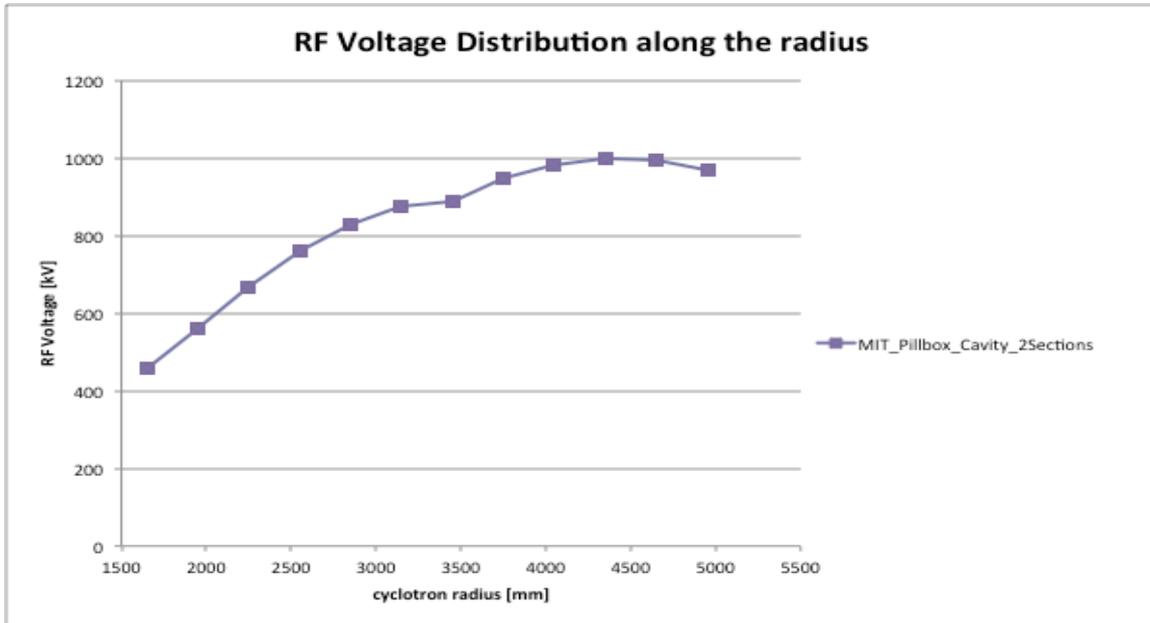

*Figure 9.5: SGC RF voltage distribution on the acceleration gap as function of the cyclotron radius.*

Table 9.4 Simulation results for the SGC

| Resonance mode | TM010 |
|---|---|
| Resonance frequency | ~51 MHz |
| Quality factor | ~ 29000 |
| Simulated RF Power Dissipation | ~ 450 kW |
| Voltage Distribution on a gap | 450-1000 kV |
| Max Surface Current | 6 A/cm |
| Max Electric Field | 5.5 MV/m |

**9.3 RF simulation results**

Several simulations were done to fulfil the RF specifications, such as the resonant frequency, the voltage distribution and the power dissipation constraints listed above. The simulations were performed with the eigenmode module of Microwave Studio (MWS) [CST_2011], run on a Microsoft Windows Server 2003 standard x64 Edition equipped with AMD Opteron processor 248 2.19 GHz, 7.93 GB RAM.

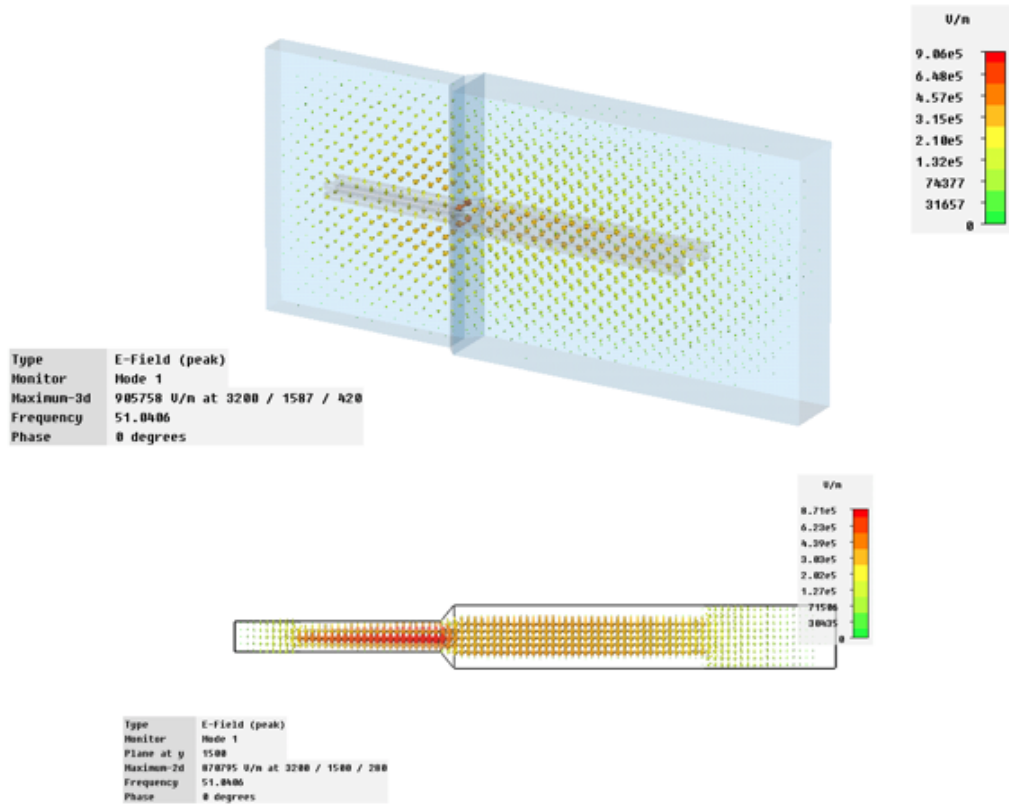

*Figure 9.6: SGC electric field on the median plane, Peak value is 5.5 MV/m.*

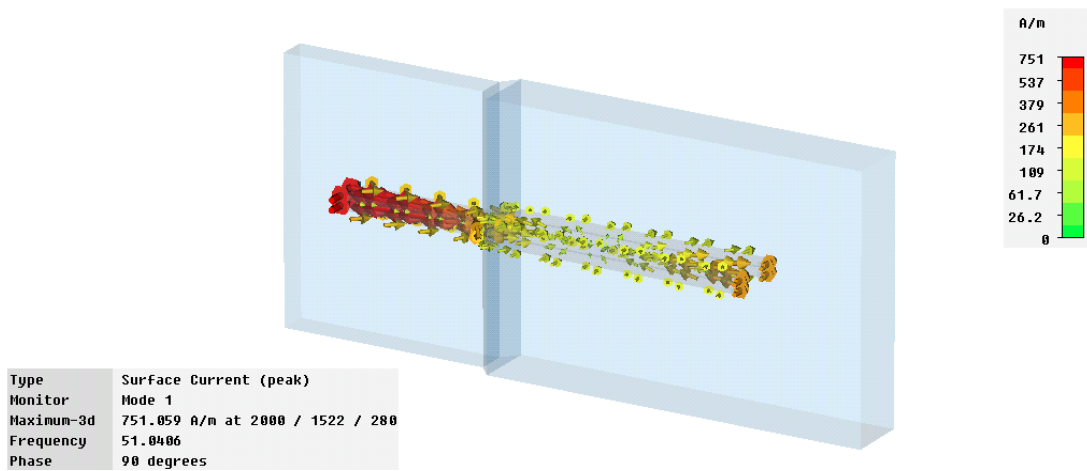

*Figure 9.7: SGC surface current distribution. Peak value is 6A/cm.*

*Single Gap Cavity (SGC)*

The SGC has a resonant frequency of 51MHz, a computed power dissipation of 165 kW, and a Q value of 29000. Table 9.4 presents the main parameters of the SGC; the voltage distribution obtained with this cavity is shown in Fig. 9.6. The computed voltage distribution is consistent with achieving the energy gain per turn required to minimize beam losses.

Fig. 9.6 shows the vector representation the electric field distribution in the median plane. The peak value is ~5.5 MV/m. The associated surface current distributions are shown in Fig. 9.7, and follows expected behavior. The maximum power dissipation occurs on the inner part of the acceleration gap, as expected. The thermal load will require attention in the design of the cooling system. Cooling will be provided by cooling channels directly TIG brazed onto the copper walls.

An inductive coupler similar to the PSI 590 MeV cavity coupler will be designed for this cavity. A trimmer for frequency tuning will be similar to that of the PSI 590 MeV cavity; the hydraulic tuning yokes will work against atmospheric pressure.

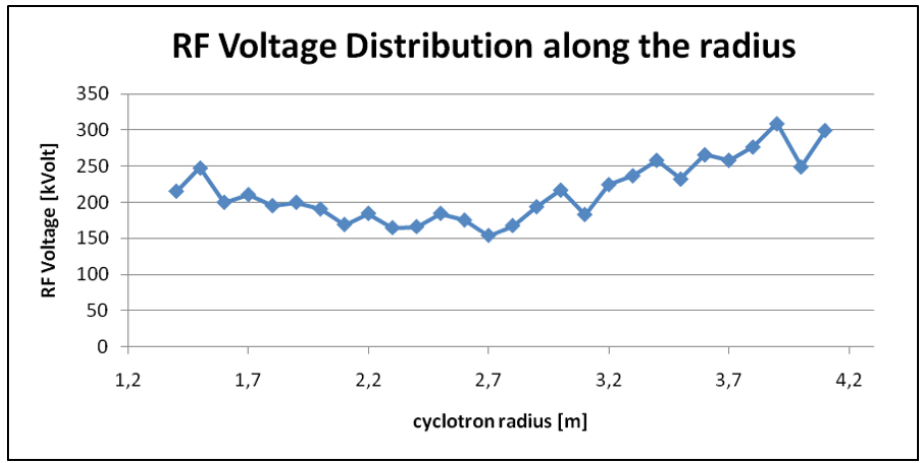

*Figure 9.8: DGC RF voltage distribution on the acceleration gap as a function of the cyclotron radius.*

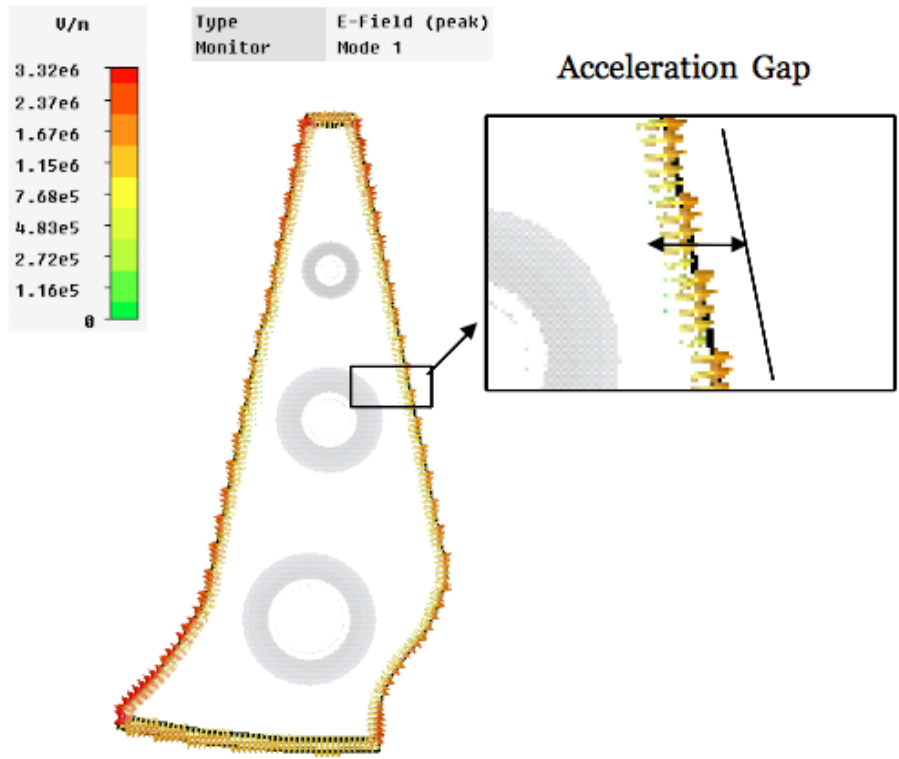

*Figure 9.9: DGC electric field on the median plane. Peak value is 20 MV/m.*

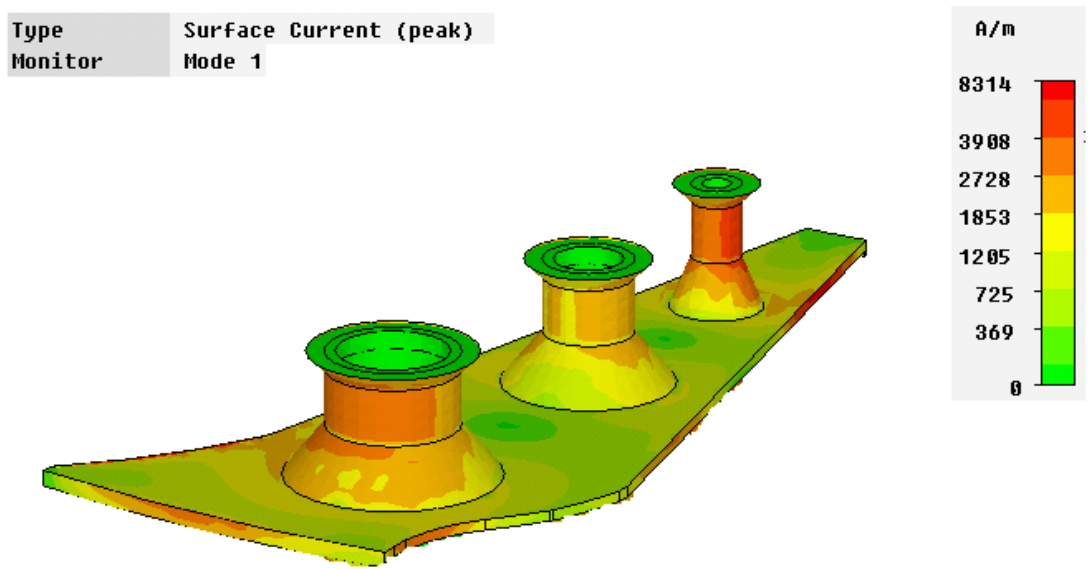

*Figure 9.10: DGC surface current distribution. Peak value is 250 A/cm.*

*Double Gap Cavity (DGC)*

The DGC operates at a resonance frequency of 51 MHz with a Q value of 13500; the computed power dissipation is 250 kW. The acceleration voltage of a DGC is the average of the voltages of the left and the right spiral acceleration gaps. The voltage distribution obtained with the cavity as simulated is shown in Fig. 9.8.

Local voltage adjustments can be made by moving the stems around their equilibrium positions. The effective voltage distribution is lower than the simulated voltage distribution and depends on the angular width Δθ of the Dee. [Magg2006].

Figure 9.9 shows the vector representation of the electric field distribution on the median plane. The peak value is ~20 MV/m. The surface current distributions are shown in Fig. 9.10. The maximum power dissipation is on the last stem and on the external part of the Dee. These parts require attention as part of the cooling system design. The cooling system will be distributed on Dee, stems, and liner in accordance with the profile of power dissipation. RF power will be fed to the cavity via an inductive coupler with area of 15 - 20cm$^2$. A frequency tuning piston on the external Dee face allows tuning the resonant frequency by 700 KHz every +/- 7 cm.

**9.4 Conclusions**

A practical RF cavity arrangement for the DAEδALUS main ring cyclotron consisting of four single-gap RF cavities and two double-gap cavities has been designed. This configuration allows accelerating the beam up to the energy requested to accomplish the DAEδALUS experiment while minimizing the beam losses due to the interaction with the residual gases. Optimization of the cavities, the engineering, and the beam loading effects require further analysis.

# 10. BEAM LOSSES

During the injection, acceleration, and extraction processes there are many sources of beam losses. The largest loss is at very low energy when the beam is injected from the ion source into the injector cyclotron. Fortunately at these energies the proton beam is not able to activate materials. Hence, the design issue is only one of thermal dissipation, which is neglected in this section. A greater concern is losses that occur once the ions have reached energies above the Coulomb barrier. Hence efficiency of beam handling at these higher energies is extremely important. So far we have addressed major design issues associated with controlling beam losses at injection and extraction. Turn separation in the injector cyclotron is adequate to ensure clean extraction from this machine. In addition, we have seen that for the SRC good solutions for both injection and extraction exist, so serious beam-loss problems are not expected in these areas.

In this section we will concentrate on two other sources of beam loss: electromagnetic dissociation (Lorentz stripping) and interactions with residual gas.

## 10.1 Lorentz force induced dissociation of $H_2^+$

Loosely-bound ions such as $H^-$, or for that matter any molecular species (including $H_2^+$), are subject to electromagnetic dissociation when traveling at high speeds in magnetic fields. Relativistic transformations show that in the rest frame of an ion, the magnetic field through which it is traveling generate a strong electric field, given in equation (10.1).

$$E = 3\beta\gamma B \text{ (MV/cm)} \qquad (10.1)$$

where B is the static magnetic field in Tesla, $\beta = v/c$ and $\gamma = 1/\sqrt{1-\beta^2}$ are the relativistic parameters of a particle with velocity v. The electric field is given in megavolts/centimeter. Table 10.1 gives some representative values of this electric field for different ion energies and magnetic fields.

Table 10.1: Electric fields seen in the rest frame of fast-moving ions travelling in magnetic fields listed. The 30 MeV case relates to commercial isotope-producing cyclotrons using $H^-$; 500 MeV is the operating point for TRIUMF, and the 800 MeV/n line represents the DAEδALUS SRC with a maximum field of 6T.

| Energy | | Electric Field (MV/cm) | | | |
|---|---|---|---|---|---|
| MeV/n | | 0.5 Tesla | 1.0 Tesla | 2.0 Tesla | 6.0 Tesla |
| 30 | | 0.38 | 0.76 | 1.53 | 4.59 |
| 500 | | 1.74 | 3.49 | 6.97 | 20.92 |
| 800 | | 2.34 | 4.68 | 9.36 | 28.08 |

Units more relevant to ionic sizes would be "volts/Angstrom," obtained by dividing the numbers in Table 10.1 by 100. One can imagine that if the electric field gradient is comparable to the binding potential of the system, distortions in the net potential will occur that can significantly lower the barrier against tunnelling, resulting in electromagnetic dissociation. This is shown schematically in Fig. 10.1.

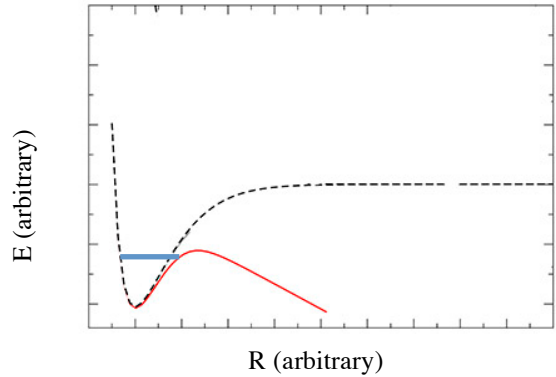

*Figure 10.1: Effect of an externally applied electric field of gradient comparable to the binding potential, showing the lowering of the barrier. This can lead to increased tunnelling and reduced stability of the ion.*

Two cases relevant to H⁻ are shown in the first two lines in Table 10.1. Commercial cyclotrons for producing isotopes routinely run (with insignificant stripping losses) at 30 MeV with maximum fields of 2T. The TRIUMF cyclotron produces H⁻ beams of 500 MeV; the maximum field in this machine is 0.5 T. Both of these cases show electric fields in the region of 1.7 MV/cm. These cases are very close to the edge, Stinson et al [Stin1969] have shown that at fields of about 2 MV/cm, H⁻ dissociation lifetime is only about 100 microseconds. At 2.5 MV/cm it is less than 1 μsec.

Evaluating this field for the case of $H_2^+$ in the DAEδALUS cyclotron, we see in Table 10.1 that the effective electric field seen by the 800 MeV/n ions in the SRC is 28 MV/cm: more than an order of magnitude higher. However, initial indications are that under proper conditions the $H_2^+$ ion will survive this field without dissociating. Fig 10.2 provides a qualitative comparison of the H⁻ and $H_2^+$ potentials, illustrating the substantial difference in depth of wells, reflected in the binding energy differences (0.75 eV for H⁻, 2.7 eV for $H_2^+$); and the larger extent of the H- potential. This latter implies that a larger field gradient will be needed to lower the potential barrier for the $H_2^+$ case, to enable tunnelling. Both of these factors contribute to the expected survival probability of the ions against Lorentz-field dissociation of the $H_2^+$ molecule.

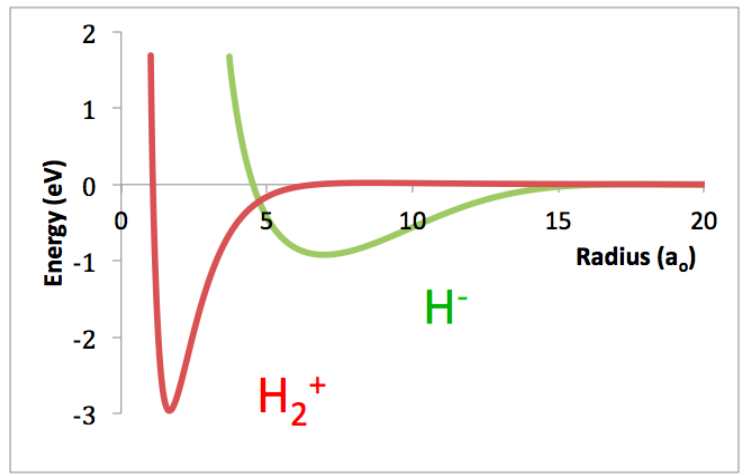

*Figure 10.2: Qualitative comparison of potential wells for H⁻ and $H_2^+$ ions. The greater depth (binding energies of 0.75 eV for H⁻ and 2.75 eV for $H_2^+$) and higher compactness of the $H_2^+$ ion's potential (well at ~2 $a_o$ (Bohr radii) for $H_2^+$ vs ~9 $a_o$ for H⁻) indicate higher probability of survival of the $H_2^+$ ion at the higher fields expected in the DAEδALUS SCR.*

The $H_2^+$ survival question, however, is complicated by the large number of stable, and very long-lived vibrational states of the $H_2^+$ ion. Because of the differences in the interatomic distances for the hydrogen molecule (0.87 A°) and the $H_2^+$ ion (~0.95 A°), the simple process of removing an electron from the neutral hydrogen molecule excites vibrational modes in the resulting ion. Calculations indicate existence of 19 stable vibrational states, and because $H_2^+$ has no dipole moment these states have very long lifetimes (> ~$10^6$ sec) [Hus_1988].

The population distribution of vibrational states [Sen_1987] is expected to follow a Franck-Condon distribution, probably independent of the source producing the ions. With a peak at $v = 2$, approximately 85% are at $v \leq 7$. Still, ~ 15% are at states higher than $v = 7$.

Fig. 10.3 shows calculations performed by J.R. Hiskes [Hisk1961] of ionic lifetimes plotted for electric fields (x axis) and binding energy (y axis) for different vibrational states of the $H_2^+$ molecule.

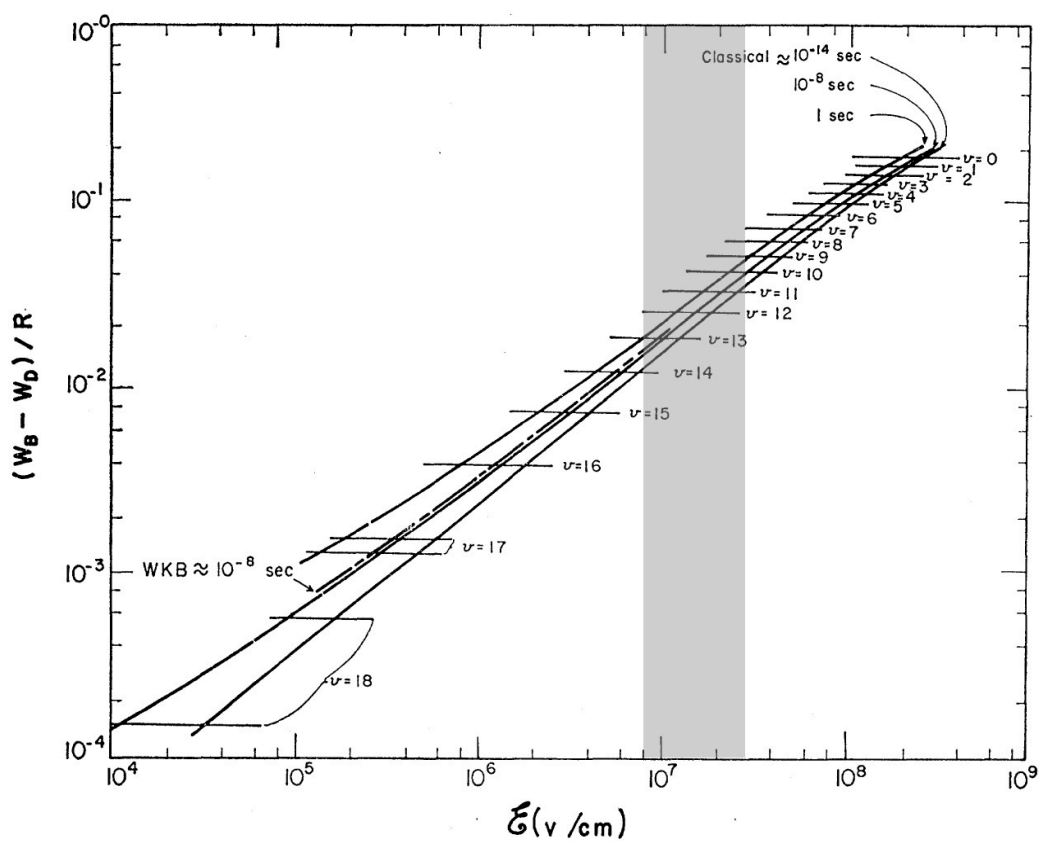

*Figure 10.3: Binding energy measured from the unperturbed dissociation limit vs electric field for the various vibrational states of the $H_2^+$ ion and for J=0 (rotational effects not considered). The intersection of the horizontal line with the curve marked "classical" determines the electric field necessary to dissociate the ion in $10^{-14}$ sec. The diagonal lines marked $10^{-8}$ sec and 1 sec determine the fields necessary for dissociation in these times, respectively. The two horizontal lines for v=18 and v=17 indicate the range of uncertainty in these calculations. The results of the WKB calculations are also indicated. [Hisk1961] (Added shaded band corresponds to Lorentz fields experienced by $H_2^+$ ions in the DAEδALUS SRC.)*

These calculations indicate that the binding energy of any state above $v \approx 7$ will dissociate in fields lower than the maxima to be encountered in the SRC, which, unfortunately will yield fragments at energies above the Coulomb barrier, so will lead to activation.

However, Hermann and Pac'ak [Herm1977], and Hus et al [Hus_1988] found that mixing helium or neon gas in the ion source will remove higher vibrational states through collision-induced dissociation reactions:

$$H_2^+(v) + He \rightarrow HeH^+ + H \qquad (10.2)$$

$$H_2^+(v) + Ne \rightarrow NeH^+ + H \qquad (10.3)$$

which were found to be exothermic for $v > 2$ and $v > 1$ respectively. If the plasma in the ion source has sufficient time to equilibrate, then only states $v = 0, 1, 2$ (for helium) and $v = 0, 1$ (for neon) would survive to be extracted and form the beam accelerated. The total current from the ion source will be reduced, but ions not extracted would have been lost at high fields in the cyclotron.

If the calculations leading to the lifetimes in Fig. 10.3 are accurate, it would appear that the $v = 2, 1, 0$ states will not be affected by electric fields below 100 MV/cm, well above the fields that will be experienced at the top energy in the SRC.

It is clear, though, that modern numerical calculations which do not need to use the approximations of Hiskes, as well as experimental tests, should be conducted to verify that Lorentz field dissociation will not be a problem for the beams we expect to accelerate. Both activities are planned for the near future.

## 10.2 INTERACTION WITH RESIDUAL GAS

Due to the interactions with the residual gas, ions can lose their orbital electron as they travel along the acceleration path. The fraction of beam particles which survives acceleration is given by [Betz1972]:

$$T = N/N_0 = \exp(-3.35 \cdot 10^{16} \int \sigma_L(E) \, P \, dL) \qquad (10.4)$$

$$\sigma_L(E) \approx 4\pi a_0^2 \, (v_0/v)^2 \, (Z_t^2 + Z_t)/Z_i \qquad (10.5)$$

where P is the pressure (torr) ($3.35 \cdot 10^{16}$ is the number of molecules/cm$^3$ in one torr), L is the path length in cm, and $\sigma_L(E)$ is the cross section of electron loss. $v$ is the ion velocity, while $v_0$ and $a_0$ are the characteristic Bohr velocity and radius. $Z_t$ and $Z_i$ are the atomic number of the residual gas and of the incident ion respectively. While this formula is in good agreement with experimental data, its accuracy is not demonstrated for energies higher than 100 MeV/n.

We evaluated the expected beam losses along the acceleration path inside the DAEδALUS SRC taking into account the energy gain per turn and of the length of the trajectory at each turn. Similarly we evaluate the beam losses along the acceleration path for the TRIUMF cyclotron. The comparison with the beam losses at TRIUMF is well founded because at TRIUMF the accelerated particle is H$^-$, a proton and 2 electrons. Moreover the second electron of the ion H$^-$ has binding energy 4 times less than the electron in the H$_2^+$, so we expect that the probability to lose the electron during the interaction with the residual gases for H$^-$ should be higher than for H$_2^+$. The results of expected beam power lost along the whole acceleration path for the SRC and for the TRIUMF cyclotron are shown in Fig. 10.4. The losses for the SRC are evaluated for two different operating vacuum values and assuming an average beam current of 2 mA $H_2^+$, which means a delivered beam power of 3.2 MW. The relevant parameters for the TRIUMF cyclotron simulation and for the SRC are shown in Table 10.2. TRIUMF cyclotron the parameters are consistent with recent years' experience. The beam power losses for the SRC are of 930 and 1800 W for an operating vacuum of $1\cdot10^{-8}$ and $2\cdot10^{-8}$ mbar respectively.

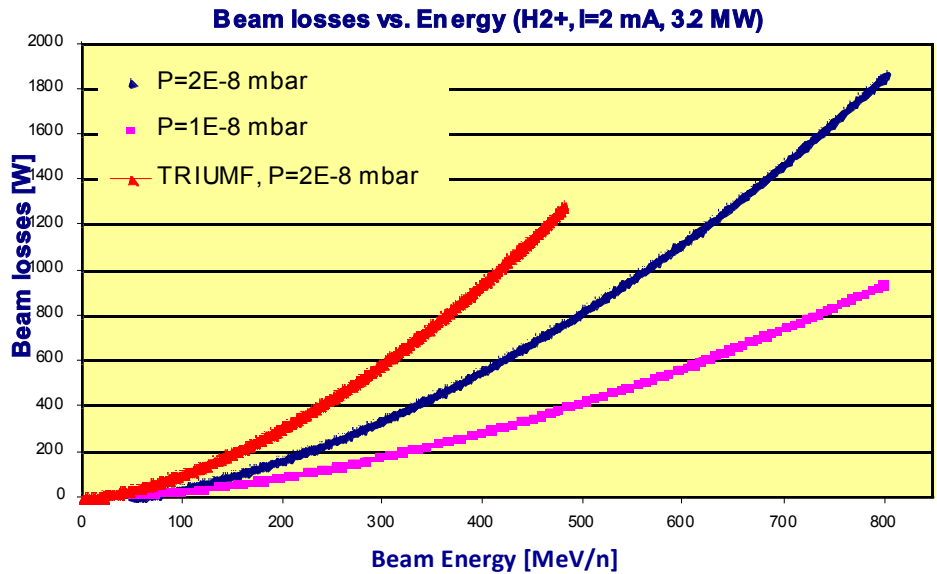

*Figure 10.4: Beam power losses vs. Beam energy. The energy gain per turn used in the simulation of the SRC- DAEδALUS is 2.4 MeV/n per turn*

Table 10.2: Beam losses due to interactions with residual gases, along the acceleration path.

|  | $E_{max}$ MeV | $\Delta E/\Delta n$ MeV | $R_{ex}$ m | $<I>$ mA | Vac. torr | $I_{loss}$ % | $I_{loss}$ μA | $P_{loss}$ W |
|---|---|---|---|---|---|---|---|---|
| TRIUMF | 480 | 0.192 | 7.6 | 0.3 | $2\ 10^{-8}$ | 2.4 | 7.2 | 1290 |
| SRC $H_2^+$ | 800 | 2.4 | 4.9 | 2 | $2\ 10^{-8}$ | 0.124 | 2.5 | 1867 |
| SRC $H_2^+$ | 800 | 2.4 | 4.9 | 2 | $1\ 10^{-8}$ | 0.062 | 1.25 | 934 |

The beam losses for the SRC were evaluated assuming that the RF cavities are able to supply an energy gain of 4.8 MeV per turn (4 PSI-like cavities plus 2 double-gap cavities). The beam power losses in the TRIUMF cyclotron are ~1290 W. The values in Table 10.2 are probably not that accurate, due to the approximations in eq. 10.4 and 10.5, but they are useful to compare the estimated losses for the SRC and TRIUMF cyclotrons. From Table 10.2, the expected beam loss percentage for the SRC cyclotron should be 20 times lower than in the TRIUMF cyclotron. In the case of the SRC, with a working vacuum of 10 nTorr and with a beam power of 3.2 MW, the beam losses are 30% lower than in the TRIUMF cyclotron. At this level they match the experience at PSI. The present evaluation considers only the beam losses due to interaction with the residual gases and neglects the losses due to electromagnetic stripping. This latter source of beam loss is more serious for the $H^-$ than for $H_2^+$. Therefore, we expect that the SRC delivering up to 3.2 MW of beam power should have beam losses at least 38% lower than the present TRIUMF cyclotron. Despite the fact that the value of the beam losses obtained by the calculation seems too high, the comparison with TRIUMF cyclotron shows that probably the real value should be 5 times lower.

We must remember that particles will be lost along the entire acceleration path and that more than 50% of the power lost is contributed by protons with energy exceeding 400 MeV. At that energy the proton range in iron is greater than 150 mm; therefore the power will be dissipated in a large volume of iron. This means that we do not have any risk of thermal hot spots due to these losses. These beam losses could be reduced if a better vacuum and/or higher energy gain per turn can be achieved. The proposed SRC cyclotron is smaller than the TRIUMF machine. Therefore, achieving a better vacuum seems possible. Moreover, a better operating vacuum is useful with respect to increasing the reliability of the RF cavities.

## 11. CONCLUSION

Accelerators for the DAE$\delta$ALUS experiment must guarantee a high level of reliability, ease of operation, as well as highly efficient conversion from electrical to beam power. This report describes an initial design study of an accelerator complex that delivers megawatt level $H_2^+$ beams at 800 MeV/n as required by the DAE$\delta$ALUS experiment.

The design presented here is feasible with the present technology. The sector magnet designs can produce a maximum average magnetic field of 1.9T with the right slope versus the radius throughout the acceleration process. Further optimizations of the spiral shape of the sector and of the coils are needed to assure a better vertical focusing and lower magnetic forces. The shape of the sectors and coils we described here allows the installation of PSI like RF cavities, which are reliable and able to achieve a maximum voltage of 1MV. The use of these cavities will reduce the number of turns in the ring cyclotron and the beam losses due the interaction with the residual gases. A most critical point needing detailed study is the design of the vacuum system capable of maintaining value of $5 \cdot 10^{-9}$ mbar required for the vacuum inside the acceleration chamber necessary to deliver a power of 6 MW with a single cyclotron complex at the site placed at 20 km. Despite the good vacuum level achieved in the TRIUMF cyclotron, the use of RF cavities to be operated at high voltage and high power, like the PSI ones, could limit this goal if not properly designed.

Accelerating $H_2^+$ ions – that can be extracted via stripping – has the potential of opening new thresholds of beam power at energies approaching 1 GeV in a very cost-effective package. Some physics questions still need to be answered related to the vibrational state population of the beam emerging from ion sources, and methods developed for quenching those states too weakly bound to survive Lorentz stripping in the high fields of the SCR. Refined calculations are being performed, and the seeds of the required experimental program are being developed. We do not expect that this will be, in the long run, a serious problem.

On several parallel fronts, we are proceeding with a well-defined plan for advancing designs, prototyping, testing and costing of critical components for this stimulating project.